\documentclass[fleqn,usenatbib]{mnras}
\usepackage{newtxtext,newtxmath}
\usepackage[T1]{fontenc}

\DeclareRobustCommand{\VAN}[3]{#2}
\let\VANthebibliography\thebibliography
\def\thebibliography{\DeclareRobustCommand{\VAN}[3]{##3}\VANthebibliography}

\usepackage{graphicx}	% Including figure files
\usepackage{amsmath}	% Advanced maths commands
\usepackage{xfrac,nicefrac}
\usepackage{subcaption}
\usepackage{enumitem}

\defcitealias{Sabhahit2022}{S22}
\title[VMS Mass-loss framework for low $Z$]{Very Massive Stars and Pair-Instability Supernovae: Mass-loss Framework for low Metallicity}

\author[G N Sabhahit et al.]{
Gautham N. Sabhahit$^{1}$\thanks{E-mail: gautham.sabhahit@armagh.ac.uk},
Jorick S. Vink$^{1}$,
Andreas A.C. Sander$^{1,2}$,
Erin R. Higgins$^{1}$\\
$^{1}$Armagh Observatory and Planetarium, College Hill, Armagh BT61 9DG, N. Ireland\\
$^{2}$Zentrum f{\"u}r Astronomie der Universit{\"a}t Heidelberg,
Astronomisches Rechen-Institut, M{\"o}nchhofstr. 12-14, 69120
Heidelberg, Germany}

\date{Accepted XXX. Received YYY}
\pubyear{2023}

\begin{document}
\label{firstpage}
\pagerange{\pageref{firstpage}--\pageref{lastpage}}
\maketitle

% Abstract of the paper
\begin{abstract}
Very massive stars (VMS) up to 200-300\,$M_\odot$ have been found in the Local Universe. If they would lose little mass they produce intermediate-mass black holes or pair-instability supernovae (PISNe). Until now, VMS modellers have extrapolated mass-loss vs. metallicity ($Z$) exponents from optically-thin winds, resulting in a range of PISN thresholds that might be unrealistically high in $Z$, as VMS develop optically-thick winds. We utilize the transition mass-loss rate of \citet{Vink2012} that accurately predicts mass-loss rates of Of/WNh ("slash") stars that characterize the morphological transition from absorption-dominated O-type spectra to emission-dominated WNh spectra. We develop a wind efficiency framework, where optically thin winds transition to enhanced winds, enabling us to study VMS evolution at high redshift where individual stars cannot be resolved. We present a MESA grid covering $Z_\odot/2$ to $Z_\odot/100$. VMS above the transition evolve towards lower luminosity, skipping the cool supergiant phase but directly forming pure He stars at the end of hydrogen burning. Below the transition, VMS evolve as cooler luminous blue variables (LBVs) or yellow hypergiants (YHGs), naturally approaching the Eddington limit. Strong winds in this YHG/LBV regime -- combined with a degeneracy in luminosity -- result in a mass-loss runaway where a decrease in mass increases wind mass loss. Our models indicate an order-of-magnitude lower threshold than usually assumed, at $Z_\odot/20$ due to our mass-loss runaway. While future work on LBV mass loss could affect the PISN threshold, our framework will be critical for establishing definitive answers on the PISN threshold and galactic chemical evolution modelling.
\end{abstract}

% Select between one and six entries from the list of approved keywords.
% Don't make up new ones.

%%%%%%%%%%%%%%%%%%%%%%%%%%%%%%%%%%%%%%%%%%%%%%%%%%

%%%%%%%%%%%%%%%%% BODY OF PAPER %%%%%%%%%%%%%%%%%%

\section{Introduction}

Observations in the last decade identified massive stars with luminosities high enough to comfortably supersede upper-mass limit of 150 $M_\odot$ \citep{Crowther2010,Best2014,Mart15}. With these local observations as well as the detection of such very massive stars (VMS) in star-forming galaxies \citep[e.g.][]{Wofford2014, Smith2016, Saxena2020}, it is vital to understand the evolution of VMS and the relevant physics affecting them. Born with an initial mass above 100 $M_\odot$ \citep{Vink2015}, VMS are thought to end their lives either as pair-instability supernovae (PISNe) leaving behind no compact remnant \citep{FH1964, Barkat1967, Woosley2002, Renzo2020}, or to directly collapse into a black hole (BH).
Intermediate to these two options, one may possibly witness pulsational pair-instability SNe (PPISN) where the star undergoes pulsations resulting in multiple mass-loss episodes \citep{Woos15,Marchant2019,Farmer2019}. 
VMS likely played a key role at high-redshift and possibly in helping re-ionise the early Universe.  
At low metallicity ($Z$) they probably ended their lives as intermediate-mass BHs \citep{Belkus2007, Yung2008}, which could bridge the gap between stellar mass and supermassive BHs. 

While there have been many suggestions for the existence of PPISNe \citep{Woos07,Gomez19,Leung20,Woos22,Wang22}, claims for the existence of full-fledged PISNe \citep{Gal09} have been relatively rare in the Local Universe.
Given the existence of VMS up to 200-300$M_\odot$ \citep{Best2020,Kalari22,Brands22}, it is hard to understand the lack of PISN explosions if stars lose no mass. 
In reality, local VMS probably lose rather large amounts of mass in stellar winds, but this wind evaporation effect is likely to be weaker at lower metallicity $Z$ \citep{Vink2018}. In other words, one might expect a threshold $Z$ below which PISNe could occur \citep{Heger2003}, and above which they might be absent.  \citet{Langer2007} used non-rotating VMS models and predicted a threshold of about $Z_\odot/3$, while VMS models from \citet{Spera2017} suggest a threshold of $\approx Z_\odot/2$. While this number has varied between a third to half solar metallicity in the past, the key to understanding this threshold with cosmic time is the $Z$-dependent mass-loss of VMS.

While canonical massive O-type stars can be reasonably explained with the classical line-driven wind theory by \citet{CAK1975}, the stellar winds for VMS have been predicted \citep{Vink2011} and observed \citep{Best2014} to be boosted above a certain transition point.  The spectral appearance of these VMS closely resemble those of emission-line dominated Wolf-Rayet (WR) stars of the nitrogen (N) sequence (WN) with hydrogen (H). These WNh stars are close to the Eddington limit \citep[e.g.][]{GH2008} and likely still core-H burning objects \citet{Massey1998, Dekoter1998}. As such WNh stars are typically found in young massive clusters, they are interpreted as part of the spectral type sequence continuing on top of classical O stars ("O stars on steroids") \citep{CW2011}. 

Although we cannot resolve individual stars in the high-redshift Universe, some stellar models have probed the evolution and fates of VMS in the early Universe \citep[e.g.][]{Chen2015, Spera2017,Volpato2023}, with VMS mass-loss rates scaled downwards with $Z$ and $\Gamma$, but the vast majority of studies on VMS evolution have been computed with mass-loss metallicity relations that were linearly extrapolated to low $Z$ \citep[e.g.][]{Yusof2013, Kohler2015}. Theoretical studies indicate that the scaling of the mass-loss rate ($\dot{M}$) with $Z$ can deviate from a power-law \citep[e.g.][]{Vink2001, Kudritzki2002, SV2020}, even if winds remain optically thin.

For VMS, there is an additional complexity due to their winds becoming optically thick. \cite{Sabhahit2022} (henceforth S22) implemented a new mass-loss prescription for high $Z$, where VMS winds can be calibrated against observations using the transition mass-loss rate concept of \cite{Vink2012}, who characterized the transition from classical O stars to WNh stars using wind efficiency numbers and sonic point optical depths. The concept of transition mass loss can be used to accurately predict mass-loss rates of the transition Of/WNh (slash) stars \citep[see also][]{Sabhahit2022}. In this paper, we further develop the theoretical framework of transition mass loss to study the evolution of VMS also in low-$Z$ environments, which are critical for high redshift galaxies, but for which we cannot observe individual transition-type stars.

The paper is organized as follows. In Sec. \ref{sec: winds_VMS}, we discuss the concept of transition mass loss and use stellar atmosphere models to parameterize the wind efficiency number at the transition point. In Sec. \ref{sec: transition_prop_predict}, we perform a simple exercise to predict properties of the `slash' stars in the Arches cluster and the 30 Dor. In Sec. \ref{sec: vms recipe Z}, we detail the framework for a new VMS mass loss approach in the stellar evolution code MESA that can be extended to investigate the evolution of VMS at low $Z$ (Sec. \ref{sec: mesa}). The implications of enhanced mass loss on PISNe explosions at low $Z$ are discussed in Sec. \ref{sec: PISN}. Finally, Sec. \ref{sec: conclusions} concludes the paper with a summary of the results and conclusions.

\section{Enhanced winds of Very Massive Stars}
\label{sec: winds_VMS}

With the effects of mass loss on massive star evolution becoming increasingly important towards the upper end of the initial mass spectrum, it is crucial to adopt optically-thick wind physics in VMS evolution models. For over a decade, we have known that VMS wind models from Monte Carlo (MC) calculations predict a higher absolute mass-loss rate in their winds, i.e., they are \textit{enhanced}, compared to their classical massive star counterparts for the same luminosity \citep{Vink2006}.
These MC calculations were further improved to include dynamical consistency \citep{MV2008}. \citet{Vink2011} used this technique to investigate the mass-loss properties of VMS up to 300 $M_\odot$. They demonstrate the existence of a \textit{kink} in the mass loss as a function of the electron scattering Eddington parameter $\Gamma_\mathrm{e}$. The quantity $\Gamma_\mathrm{e}$ is given by the ratio of Thomson acceleration $a_\mathrm{thom}$ and Newtonian gravity $g$:
\begin{equation}
\begin{array}{c@{\qquad}c}
\Gamma_\mathrm{e} = \dfrac{a_\mathrm{Thom}}{g} = \dfrac{\varkappa_\mathrm{Thom}L}{4\pi G cM} \approx 10^{-4.813}(1+X_\mathrm{s})\dfrac{L}{M}
\end{array}
\label{eq: gamma_e}
\end{equation}
The last equality is true at high enough temperatures where the gas is assumed to be fully ionized and the electron scattering opacity can be approximated as $\varkappa_\mathrm{Thom} \approx 0.2(1+X_\mathrm{s})$ $\mathrm{cm}^2\; \mathrm{g}^{-1}$. For a constant surface hydrogen abundance $X_\mathrm{s}$, $\Gamma_\mathrm{e}$ becomes proportional to the luminosity-to-mass ratio. Above this kink, \citet{Vink2011} find their mass-loss rates to scale steeply with $\Gamma_\mathrm{e}$, effectively enhancing the winds compared to classical massive stars below the kink (see their Fig. 5). 

On the observational side, \citet{Best2014} performed spectral analysis of O and WNh stars in the 30 Dor cluster using the non-LTE radiative transfer code CMFGEN to obtain both stellar and wind parameters. They too find a kink in the observed mass-loss rates, above which mass loss scales steeply with $\Gamma_\mathrm{e}$. 

\citet{Vink2011} connect this enhancement in the mass loss above the kink to the wind efficiency number given by 
\begin{equation}
\begin{array}{c@{\qquad}c}
\eta = \dfrac{\dot{M} \varv_\infty}{(L/c)}
\end{array}
\label{eq: eta}
\end{equation}
surpassing the single scattering limit of $\eta = 1$. In the Arches cluster, one can evaluate the average wind efficiency number of the two Of/WNh stars assuming a micro-clumping factor $D_\mathrm{cl}$ = 10 (or volume filling factor $f_\mathrm{V} = 0.1$). The average $\eta$ obtained is lower compared to the single scattering limit of $\eta = 1$ predicted in the MC calculations. However, we note that the MC models over-predict the terminal velocities compared to the empirically determined values, thus probably under-predicting the absolute mass-loss rates. Consequently, the location of the kink would then shift to lower values of $\eta$. Since we base our new implementation on wind efficiency numbers, it is important to quantify the absolute value of $\eta$ at the kink as well as understand its variation with stellar parameters.

\citet{Vink2011} also studied the behaviour of the predicted He \textsc{ii} 4686 \AA \; line in the optical for three increasing values of $\Gamma_\mathrm{e}$. The He \textsc{ii} 4686 \AA \; line is a recombination line, meaning its equivalent width is sensitive to the density in the wind and \citet{Vink2011} demonstrated that the equivalent widths of the predicted He \textsc{ii} lines indeed increase sharply with $\Gamma_\mathrm{e}$. Models above the predicted $\Gamma_\mathrm{e, kink}$ showed very strong and broad He \textsc{ii} emission lines that are characteristic of the stars in the WNh sequence.  They concluded that the observed spectral morphological transition from O to the WNh sequence corresponds to a switch from optically thin to thick wind driving, with the kink corresponding to the transition `slash' stars.

The following sub-sections are organized as follows. In Sec. \ref{sec: transition mass loss}, we discuss the transition mass loss and how one can empirically determine it in a young massive cluster.  In Sec. \ref{sec: f_variation_with param} we discuss how the wind efficiency numbers might vary with stellar parameters when the sonic point optical depth crosses unity, and parameterize the transition using stellar atmosphere models. Later in Sec. \ref{sec: vms recipe Z}  we briefly mention our previous attempt at implementing a VMS mass loss in \citetalias{Sabhahit2022}, and further delve into the details of a new VMS mass-loss framework based on the theory of transition mass loss which can be extended to low $Z$.

\subsection{Transition mass loss}
\label{sec: transition mass loss}

\begin{figure}
    \includegraphics[width = \columnwidth]{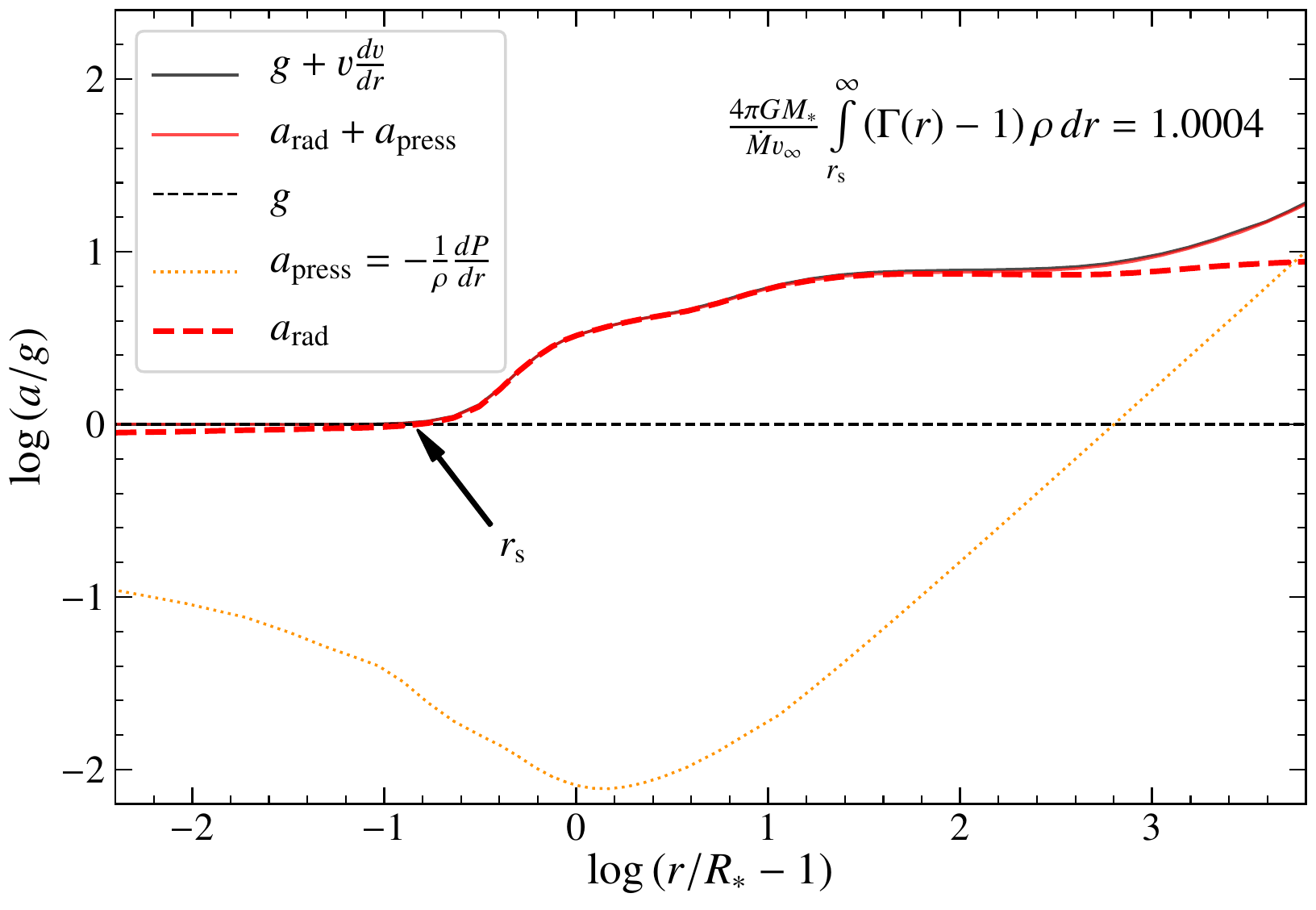}
    \caption{Acceleration stratification for a hydro-dynamically consistent model from PoWR. The model parameters are log($L/L_\odot$) = 6.39, $M/M_\odot = 97$, effective temperature at $T_\mathrm{eff}(R_{\tau_\mathrm{Ross} = 2/3}) = 42.7$ kK, $X_\mathrm{s} = 0.7$ and $Z_\mathrm{Fe} = 1.6 \times 10^{-3}$. }
    \label{fig:Powr_check_eq}
\end{figure}

We closely follow the arguments used in \citet{Vink2012} to derive the transition mass loss and set up the relevant formulas to extend our study of the transition point to lower $Z$. Using the hydro-dynamic equation of motion, \citet{Vink2012} derived a model-independent way to characterize the transition in the sonic point conditions from optically-thin to optically-thick wind driving.   Assuming stationary ($d\varv/dt = 0$) winds, the hydro-dynamic equation of motion can be written as 
\begin{equation}
\begin{array}{c@{\qquad}c}
\varv\dfrac{d\varv}{dr} = -\dfrac{1}{\rho}\dfrac{dP_\mathrm{gas}}{dr} - \dfrac{GM_\star}{r^2} + a_\mathrm{rad}(r)
\end{array}
\label{Eq: Hydro_eqn}
\end{equation}
The above equation connects the advective acceleration of motion ($\varv d\varv/dr$) to the different forces acting in the wind such as Newtonian gravity, gas and radiative pressure forces. Following \citet[Eq.\,7.5]{LC1999}, the hydro-dynamic equation is integrated from the inner boundary $R_\star$ to $\infty$, through the sonic point $r_\mathrm{s}$ where the wind velocity equals the sonic velocity (cf. Eq. \ref{eq: sonic_velo}) 
\begin{equation}
\begin{split}
& \displaystyle\int_{R_\star}^{\infty} 4\pi r^{2}\rho\varv\dfrac{\mathrm{d}\varv}{\mathrm{d}r}dr + \displaystyle\int_{R_\star}^{r_\mathrm{s}} \Biggr[\dfrac{1}{\rho}\dfrac{dP_\mathrm{gas}}{dr} + \dfrac{GM_\star}{r^2}\Biggr]dm \\ & + \displaystyle\int_{r_\mathrm{s}}^{\infty} \dfrac{1}{\rho}\dfrac{dP_\mathrm{gas}}{dr}dm + \displaystyle\int_{r_\mathrm{s}}^{\infty} 4\pi r^{2}\rho\Biggr[\dfrac{GM_\star}{r^2} - a_\mathrm{rad}(r)\Biggr]dr = 0
\end{split}
\label{Eq: Hydro_eqn_first_form}
\end{equation}
This integral form can be further simplified by realizing that in the sub-sonic region of the wind ($r<r_\mathrm{s}$), hydro-static equilibrium is a good approximation and the contribution from the second term is negligible. Likewise, the neglection of the gas pressure forces is a good approximation above the sonic point ($r>r_\mathrm{s}$) and allows us to omit the third term. The first term is the momentum in the wind $\dot{M}\varv_\infty$. The radiative pressure term in the hydro-dynamic equation is expressed in terms of the gravity using the Eddington parameter $\Gamma(r)$:
\begin{equation}
\begin{array}{c@{\qquad}c}
\Gamma(r) = \dfrac{a_\mathrm{rad}(r)}{g} = \dfrac{1}{gc}\displaystyle\int_{0}^{\infty} \varkappa_\nu F_\nu d\nu = \dfrac{\varkappa_{F}(r)L}{4\pi G cM_\star} 
\end{array}
\label{eq: Gamma}
\end{equation}
where $\varkappa_{F}$ is the flux-weighted mean opacity (in $\mathrm{cm}^2 \mathrm{g}^{-1}$) calculated in the co-moving frame which accounts for the Doppler shift in the frequencies due to non-zero wind velocities. In contrast to the electron scattering Eddington parameter in Eq. (\ref{eq: gamma_e}), the total Eddington parameter $\Gamma(r)$ is a function of distance $r$ through its dependence on $\varkappa_{F}(r)$. Using these simplifications, Eq. (\ref{Eq: Hydro_eqn_first_form}) reduces to
\begin{equation}
\begin{split}
& \dot{M}\varv_\infty  =  4\pi G M_\star\displaystyle\int_{r_\mathrm{s}}^{\infty} (\Gamma(r)-1)\rho dr
\end{split}
\label{Eq: Hydro_eqn_sec_form}
\end{equation}
In Fig. \ref{fig:Powr_check_eq}, we test the validity of the assumptions made in deriving Eq. (\ref{Eq: Hydro_eqn_sec_form}) by using a hydro-dynamically consistent atmosphere model from the Potsdam Wolf-Rayet (PoWR) code (for details of model parameters see Sec. \ref{sec: f_variation_with param}). The plot shows different accelerations in the atmosphere normalized to gravity. The hydro-dynamical equation of motion is satisfied throughout the atmosphere, shown by the super-imposed gravity + inertia term and the gas + radiative pressure term (black and red solid lines). The red dashed line is the total radiative acceleration $a_\mathrm{rad}(r)$ with contributions from both continuum processes (electron scattering, bound-free and free-free) and line scattering. We perform the integral of $(\Gamma(r) - 1)$ weighted with density $\rho$ from $r_\mathrm{s}$ to the outer boundary of our model. In the top right corner, we give the value of this integral multiplied by $4\pi G M_\star/\dot{M}\varv_\infty$, which is very close to unity. 

Using the total Eddington parameter and the flux-weighted mean optical depth at the sonic point $\tau_{F,\mathrm{s}} = \int_{r_\mathrm{s}}^{\infty} \varkappa_{F}\rho dr$, Eq. (\ref{Eq: Hydro_eqn_sec_form}) can be further simplified to
\begin{equation}
\begin{array}{c@{\qquad}c}
\dot{M}\varv_\infty = \dfrac{L}{c}\Big(\dfrac{\Gamma-1}{\Gamma}\Big)_\mathrm{avg}\tau_{F,\mathrm{s}}
\end{array}
\label{eq: transition_mdot_first_form}
\end{equation}
where a suitable radial average of ($\Gamma-1)/\Gamma$ is performed in the supersonic region of the wind. In the limiting case of $\Gamma \gg 1$ above the sonic point, Eq. (\ref{eq: transition_mdot_first_form}) gives a simple relation at transition when $\tau_{F,\mathrm{s}}$  becomes unity
\begin{equation}
\begin{array}{c@{\qquad}c}
\eta = \tau_{F,\mathrm{s}}  = 1 \;\;\;\;\;\; \mathrm{or} \;\;\;\; \dot{M}_\mathrm{trans} = \dfrac{L_\mathrm{trans}/c}{\varv_\mathrm{\infty,trans}} 
\end{array}
\label{eq: transition_mdot_sec_form}
\end{equation}
\citet{Vink2012} tested the validity of the assumption of $\Gamma \gg 1$ using the PoWR code \citep{Grafener2002, HG2003, Sander2015} and determined a correction factor.
\begin{equation}
\begin{array}{c@{\qquad}c}
\eta = f\cdot \tau_{F,\mathrm{s}} = f \;\;\;\;\;\; \mathrm{or} \;\;\;\; \dot{M}_\mathrm{trans} = f\dfrac{L_\mathrm{trans}/c}{\varv_\mathrm{\infty,trans}} 
\end{array}
\label{eq: transition_mdot_third_form}
\end{equation}
They find $f \approx 0.6$ at the transition point from their Galactic models (see Sec. \ref{sec: f_variation_with param} for details), i.e., the flux-weighted mean optical depth at the sonic point crosses unity for $\eta \approx 0.6$ at Galactic metallicity.  Given a cluster of O and WNh stars with empirically determined luminosities $L_\mathrm{trans}$ and terminal velocities $\varv_\mathrm{\infty, trans}$ of the `slash' stars, one can use Eq. (\ref{eq: transition_mdot_third_form}) to obtain the so-called \textit{transition mass-loss rate}. The accuracy with which one can determine the transition mass-loss rate then only depends on the uncertainties in the stellar luminosity, with terminal velocity well constrained from the ``blue" absorption edge of P Cygni lines.

The transition mass-loss rate so derived for the two `slash' objects in the Arches cluster using $f = 0.6$ evaluates to log($\dot{M}_\mathrm{trans}) \approx -5.16 \; M_\odot \; \mathrm{yr}^{-1}$. This value agrees well with empirically determined mass-loss rate of log($\dot{M}_\mathrm{emp, D = 10}) = -5.19 \; M_\odot \; \mathrm{yr}^{-1}$  for the same objects using micro-clumping factor $D_\mathrm{cl} = 10$ \citep{Martins2008}. Furthermore, the oft-used optically-thin mass-loss recipe from \citet{Vink2001} with $M = 60\; M_\odot$ gives log($\dot{M}_\mathrm{Vink01}) = -5.14 \; M_\odot \; \mathrm{yr}^{-1}$, thus correctly predicting the rates near the transition point. Agreement amongst all three independent methods of determining the mass-loss rate of the transition objects provides confidence in the theory of transition mass loss. 
%In Sec. \ref{Sec: f with Z} we show mass loss predictions from hydro-dynamically consistent stellar atmosphere models. 

\begin{table*}
\centering % used for centering table

\begin{tabular}{c c c c c c c c c c c c c} 
        \hline

         log $L_\star/L_\odot$ & $X_\mathrm{s}$ &$T_\star$[kK] & $T_\mathrm{eff}(R_{\tau_\mathrm{Ross} = 2/3})$[kK] & $Z_\mathrm{Fe}$& $M_\star/M_\odot$     & log $\dot{M}$ [$M_\odot \;\mathrm{yr}^{-1}$] & $\varv_\infty [\mathrm{km \;s^{-1}}]$ \\

         \hline\hline

         $6.2$ & 0.4  & 45 & 42.1  & 1.6$\times 10^{-3}$ & 52 & -4.99 & 1538.4 \\
         $6.39$ & 0.4  & 45 & 42.6 & 1.6$\times 10^{-3}$ & 81 &  -4.822 & 1655.1\\
         $6.7$ & 0.4  & 45 & 43.1 & 1.6$\times 10^{-3}$& 160   & -4.588 & 1900.2\\
         $6.39$ & 0.4  & 35 & 33.5 & 1.6$\times 10^{-3}$  & 98 & -4.720 & 1317.1 \\
         $6.39$ & 0.4  & 55 & 52.3 & 1.6$\times 10^{-3}$ & 74 & -4.923 & 1699.4 \\
         $6.39$ & 0.2  & 45 & 42.6  & 1.6$\times 10^{-3}$ & 71  & -4.824 & 1595.7\\
         $6.39$ & 0.3  & 45 & 42.7 & 1.6$\times 10^{-3}$ & 76  & -4.810 & 1607.6 \\
         $6.39$ & 0.5  & 45 & 42.6 & 1.6$\times 10^{-3}$  & 86 & -4.807 & 1668.4\\
         $6.39$ & 0.6  & 45 & 42.8 & 1.6$\times 10^{-3}$  & 92 & -4.914 & 1777.5 \\
         $6.39$ & 0.7  & 45 & 42.7 & 1.6$\times 10^{-3}$ & 97 &  -4.899 & 1796.3\\
         $6.39$ & 0.4  & 45 & 41.6  & 0.7$\times 10^{-3}$ & 74 &   -4.808 & 1155.9\\
         $6.39$ & 0.4  & 45 & 40.4 &  0.32$\times 10^{-3}$ & 68  & -4.789 & 820.9\\
         \hline

\end{tabular}
\begin{tabular}{c c c c c c c c c c c c c}

         $R_\mathrm{crit}/R_\odot$ & $R_{(\tau_\mathrm{Ross} = 2/3)}/R_\odot$ & $\varv_\infty/\varv_\mathrm{esc}(R_\mathrm{crit})$  & $\eta$ & $\tau_{\mathrm{Thom,s}}$  & $\tau_{F,\mathrm{s}}$& $f = \eta/\tau_{F,\mathrm{s}}$\\

         \hline\hline

         24.57 & 23.72  & 1.712 &  0.487 &  0.414 & 0.996 & 0.489 \\
         29.57 & 28.83 & 1.619 & 0.498 & 0.445 & 1.05 & 0.472 \\
         41.03 & 40.29 & 1.558 & 0.481 & 0.481 & 1.051 & 0.457\\
         48.28 & 46.51 & 1.497 & 0.493 &  0.322 & 1.021 & 0.482\\
         19.53 & 19.11 & 1.414 & 0.406 & 0.475 & 0.956 & 0.425 \\
         29.64 & 28.75 & 1.67 &  0.478 & 0.425 & 1.033 & 0.463\\
         29.44 & 28.7 & 1.62 & 0.498 & 0.44 & 1.054 & 0.472  \\
         29.54 & 28.86 & 1.583 & 0.520 & 0.439 & 1.069 & 0.487  \\
         29.67 & 28.51 & 1.635 & 0.433 & 0.368 & 0.883 & 0.490\\
         29.54 & 28.62  & 1.605 & 0.453 & 0.408 & 0.946 & 0.478 \\
         30.4 & 30.2 & 1.199 & 0.359 & 0.585 & 1.031 & 0.348\\
         32.04 & 31.53 & 0.905 & 0.267 & 0.687 & 1.01 & 0.264    \\
         \hline

\end{tabular}

\caption{Results of hydro-dynamical simulations from the stellar atmosphere code PoWR. Columns (1), (2), (3) and (5) are input parameters to test the effects of luminosity, inner boundary effective temperature $T_\star = T_\mathrm{eff}(R_\star, \tau_\mathrm{Ross,cont} = 5)$, surface H and metal content on $f$. Mass in column (6) is varied to maintain a sonic point optical depth of unity. Columns (7) and (8) provide the mass-loss rate and the terminal velocity respectively. The  wind efficiency parameter $\eta$ and $f = \eta/\tau_{F,\mathrm{s}}$ are listed in columns (12) and (15). See text for rest of the columns.} 
\label{table:powr_tau1}
\end{table*}

\subsection{Variation of $f$ with stellar parameters}
\label{sec: f_variation_with param}

How does the quantity $f$ vary with relevant stellar parameters such as luminosity and temperature? From Eq. (\ref{eq: transition_mdot_first_form}), $f$ equals the average ($\Gamma-1)/\Gamma$ in the supersonic region when $\tau_{F,\mathrm{s}} = 1$. Using $P = a(r)^2\rho$ and re-arranging the terms in the hydro-dynamic equation of motion, we get
\begin{equation}
\begin{array}{c@{\qquad}c}
\varv\Bigg(1-\dfrac{a^2}{\varv^2}\Bigg)\dfrac{d\varv}{dr} = \dfrac{GM_\star}{r^2}\Bigg(\dfrac{\varkappa_{F}(r)L}{4\pi G cM_\star}  - 1\Bigg)+ 2\dfrac{a^2}{r} - \dfrac{da^2}{dr} 
\end{array}
\label{Hydro_eqn_final_form}
\end{equation}
 where $a(r)$ is the iso-thermal sound speed corrected for a micro-turbulence term $\varv_\mathrm{mic}$,
\begin{equation}
\begin{array}{c@{\qquad}c}
a^2(r) = a^2_\mathrm{iso, sonic}(r) + \dfrac{1}{2}\varv^2_\mathrm{mic}\mathrm{.}
\end{array}
\label{eq: sonic_velo}
\end{equation}
The full expression for $\Gamma(r)$ (Eq. \ref{eq: Gamma}) is used in the hydro-dynamic equation (\ref{Hydro_eqn_final_form}) to provide a better intuition on how different stellar parameters affect the force balance. For example, a higher luminosity results in a higher radiative acceleration, thus increasing $\Gamma$ in both the sub- and super-sonic regions of the atmosphere. A higher $\Gamma$ in the sub-sonic region shifts the critical point, where $\Gamma$ crosses unity, inwards. The mass-loss rate is increased as a result of the higher radiative acceleration in the sub-sonic region. A secondary effect of this increase in radiative acceleration in the inner wind can be an effective reduction of the radiative acceleration in the super-sonic region of the wind potentially lowering the terminal velocity \citep[see numerical experiments in][]{Vink99}. Thus despite knowing the first-order effect of individual stellar parameters on the $\Gamma$ parameter, it is not straightforward to quantify them as secondary effects might influence the $\Gamma$ parameter in the super-sonic region. 

To regain the sonic point condition of $\tau_{F,\mathrm{s}} = 1$, further changes have to be made to input parameters such as the stellar mass. Since mass appears in the denominator of the $\Gamma$ parameter, increasing the mass decreases  $\Gamma$ in the sub-sonic region. The critical point shifts outwards while the mass-loss rate reduces. If this exercise of first increasing a certain parameter and then compensating by changing the mass reveals negligible variation in the super-sonic $\Gamma$, then $f$ remains almost independent of that parameter. Detailed atmosphere models will provide better insight into the variation of the $\Gamma$ parameter in the super-sonic region with different parameters.  

\citet{Vink2012} used $\beta$-law PoWR models at Galactic $Z$ to infer absolute values of $f$, and its dependence on stellar parameters. These models were calibrated based on a hydro-dynamically consistent atmosphere model \citep[of WR22 from][]{GH2008}, but using empirical information for the terminal velocity (see their Table 1). They varied both the stellar luminosity and the temperature while keeping the ratio $\varv_\infty/\varv_\mathrm{esc}$ fixed. The factor $f$ from these tests remained rather unchanged and showed no systematic variation. Furthermore, reducing the ratio $\varv_\infty/\varv_\mathrm{esc}$ regardless of the luminosity or effective temperature also reduced the value of $f$. \textit{They concluded that $f$ primarily depends on the ratio $\varv_\infty/\varv_\mathrm{esc}$ and is independent of both luminosity and temperature. }

In this section, we present stellar atmosphere models using the PoWR code which is updated to consistently solve the wind hydro-dynamics to obtain the absolute mass loss and velocity stratification simultaneously \citep[][Sabhahit et al. 2023 in prep]{Sander2017, Sander2023}. The hydro-dynamic equation of motion is also now involved in the iteration along with solving the radiative transfer and the statistical rate equations consistently.  Unlike the exercise performed in \citet{Vink2012} where they varied wind parameters such as $\varv_\infty/\varv_\mathrm{esc}$ and $\beta$ and tested its effect on the factor $f$, the hydro-dynamic implementation consistently solves for absolute mass-loss rate $\dot{M}$ and the velocity stratification in the atmosphere for a given set of input parameters.
%Although we do not take any absolute values of $f$ from these hydro-models from PoWR, these models help us gain insight as to why the ratio $\varv_\infty/\varv_\mathrm{esc}$ remains almost constant despite changes to certain parameters. With detailed atmosphere models, we can also obtain a relation connecting $\eta/\tau_{F,\mathrm{s}}$ and $\varv_\infty/\varv_\mathrm{esc}$ more suitable for VMS. 

%where 
%Furthermore, the derived values of $\dot{M}$ and $\varv_\infty/\varv_\mathrm{esc}$ at $\tau_{F,\mathrm{s}} = 1$ with the empirical results. This might hint at the mass loss or terminal velocity properties of VMS predicted by the PoWR code and inform a full parameter-space study in the future. 

In Table \ref{table:powr_tau1} we provide a list of hydro-dynamically consistent PoWR models whose flux-weighted mean optical depth is approximately unity. The inputs tested are the luminosity log$(L_\star/L_\odot)$, surface hydrogen $X_\mathrm{s}$, surface metal content $Z$ and the inner boundary effective temperature $T_\star$ defined at radius $R_\star(\tau_\mathrm{Ross, cont} = 5)$. The effective temperature at Rosseland mean optical depth of $2/3$ is also listed. The stellar mass $M_\star$ in column (6) is varied till the wind density becomes high enough to reach $\tau_{F,\mathrm{s}} \approx 1$.  In column (5) we provide the absolute iron metal mass fraction, which is a proxy to the total metallicity as we linearly scale all other metal abundances by a factor of approximately 1, 0.5, and 0.2 to obtain a metal content corresponding to our Galaxy, the Large (LMC) and Small Magellanic Clouds (SMC) respectively. All models are also enhanced in nitrogen and helium (He) with some H still left in the atmosphere, typical for the WN sequence of the `h' type. The metal mass fractions of the different elements are provided in Table \ref{table: elements} in Appendix \ref{Appendix: powr}.

The hydro-dynamic approach allows us to directly infer the value of $\eta$ at $\tau_{F,\mathrm{s}} \approx 1$. The flux-weighted mean optical depth $\tau_{F,\mathrm{s}}$ is evaluated at the critical radius $R_\mathrm{crit}$ defined where the wind velocity $\varv(r)$ crosses $a(r)$.  A constant micro-turbulent velocity of 30 km $\mathrm{s}^{-1}$ is used in our models. We take the critical radius here as it is a more meaningful parameter compared to the inner boundary radius $R_\star$. Columns (7) and (8) give the hydro-predicted mass-loss rate and the terminal velocity. The ratio of terminal-to-escape velocity $\varv_\infty/\varv_\mathrm{esc}$ is given in column (11). The escape velocity is also defined at the critical radius. 
\begin{equation}
\begin{array}{c@{\qquad}c}
\varv_\mathrm{esc} = \sqrt{\dfrac{2GM_\star}{R_\mathrm{crit}}}
\end{array}
\label{eq: v_esc}
\end{equation}
In Table \ref{table:powr_tau1}, we list two different radii -- $R_\mathrm{crit}$ and the radius where Rosseland mean optical depth equals $2/3$ (columns 9 and 10). The latter radius is relevant for stellar evolution models and usually defines their outer boundary. The difference between using $R_\mathrm{crit}$ or $R_{\tau_\mathrm{Ross} = 2/3}$ is quite small and results in only a few tens of km/s difference in the escape velocity.

We further provide the Thomson optical depth (column 13) at the critical radius, obtained by integrating the electron scattering opacity. Recently, \citet{Sen2023} applied the oft-used optical depth parameter from \citet{Langer1989}:
\begin{equation}
\begin{array}{c@{\qquad}c}
\tau_\mathrm{wind}(R) = \dfrac{\varkappa_\mathrm{e}\dot{M}}{4\pi R(\varv_\infty-\varv_\mathrm{o})}\mathrm{ln}(\varv_\infty/\varv_\mathrm{o}),\;\;\varv_\mathrm{o} = \mathrm{20\; km/s}
\end{array}
\label{eq: tau_wind_L89}
\end{equation}
to the observed Of/WNh type stars taken from \citet{Shenar2019} and \citet{Hainich2014}, and obtained $\tau_\mathrm{wind}$ values of approximately 0.2.
The formula for $\tau_\mathrm{wind}$ approximates the opacity using the electron scattering opacity and assumes a $\beta$-velocity law for the velocity stratification with $\beta = 1$. The Thomson optical depths in our hydro-models also account only for electron scattering, but use the detailed velocity law obtained from the integration.

The systematic difference between the Thomson optical depth values in our models and that reported by \citet{Sen2023} could be due to the $\beta = 1$ assumption made while evaluating $\tau_\mathrm{wind}$. Moreover, our values for $\varv(R_\mathrm{crit})$ are in the range of $25$ to $30\,\mathrm{km}\,\mathrm{s}^{-1}$ and this is slightly higher than their assumption of $\varv_\mathrm{o}$. However, this difference is negligible in its impact on Eq. (\ref{eq: tau_wind_L89}). Given our higher $\tau_\mathrm{Thom}$ values, it could be that the `slash' stars actually have slightly lower $\tau_{F,s}$ values than the unity condition we use in this work. A full parameter study in the future should provide a better handle on the value of $\tau_{F,s}$ and its association to the spectral transition. 

Our dynamically consistent PoWR models also include a micro-clumping factor $D_\mathrm{cl}$ which indicates the density increase in the clumps compared to a smooth wind and a void inter-clump region. $D_\mathrm{cl}(r)$ is considered to be a depth-dependent quantity where the increase of clumping is described by a characteristic velocity $\varv_\mathrm{cl} = 100 \; \mathrm{km/s}$ \citep{Martins2009, Sander2017}. $D_\mathrm{cl}$ is unity at the inner boundary $R_\star$ and reaches a maximum value of 10 at the outer boundary of the atmosphere defined at $R_\mathrm{max} = 10,000 \; R_\star$.

Consider the first three models in Table \ref{table:powr_tau1}. The luminosity is varied while keeping $\tau_{F,\mathrm{s}} \approx 1$. Increasing the stellar luminosity drives a denser wind with a higher mass-loss rate. At constant temperature, this also means an increase in the inner boundary radius and a reduction in escape velocity. We then increase the mass to compensate for the higher mass-loss rate. Increasing the mass also increases the escape velocity. The ratio $\varv_\infty/\varv_\mathrm{esc}$ is not affected significantly. The next two models ({cf. rows 2, 4 and 5}) test the effects of varying temperatures. Similar to the previous case, the ratio $\varv_\infty/\varv_\mathrm{esc}$ only varies slightly. Both luminosity and temperature do not significantly change $\varv_\infty/\varv_\mathrm{esc}$ while maintaining $\tau_{F,\mathrm{s}} = 1$, thus confirming the results obtained in \citet{Vink2012}. 

Now we perform additional tests to include the case of varying $X_\mathrm{s}$ (cf.  rows 2, 6, 7, 8, 9 and 10). Decreasing the amount of hydrogen in the atmosphere means a lower electron number density, which decreases the total radiative acceleration and the total particle density near the critical point. The stellar mass has to be increased to lower the outward pointing radiative acceleration to ensure $\tau_{F,\mathrm{s}} $ returns back to unity. We find the ratio $\varv_\infty/\varv_\mathrm{esc}$ to remain independent of $X_\mathrm{s}$ as well. 

Finally, we change metal opacities throughout the wind (cf. rows 2, 11 and 12). When the metal content in the sub-sonic region is lowered, the radiative acceleration is reduced. The first-order effect is a lower mass-loss rate and a faster wind. However, metal opacities are lowered in the super-sonic region as well, which is strong enough to effectively reduce the terminal velocity \citep[see also][]{Leitherer1992, SV2020, VS2021}. 
A reduction in mass to regain  $\tau_{F,\mathrm{s}} = 1$ further lowers the terminal velocity. Our models predict the ratio $\varv_\infty/\varv_\mathrm{esc}$ to vary steeply with $Z$. We do not adopt any absolute values of $f$ as predicted by these models as the comparison with observations indicate that our models might underestimate the observed values of $\varv_\infty$  (cf. Table \ref{tab: empirical_transition}). Instead, we try to find a general functional relation between $\eta$ and $\tau_{F,\mathrm{s}}$.

In addition to $\varv_\infty/\varv_\mathrm{esc}$, \citet{Vink2012} also varied the $\beta$ parameter that defines the gradient of the velocity stratification in the wind: $\varv = \varv_\infty (1-R_\star/r)^\beta$ and can have a significant influence on the spectral appearance of emission-line stars \citep[e.g.][]{Lefever2023}. Increasing the value of $\beta$ to 1.5, a value more appropriate for VMS \citep[see][]{Vink2011}, lowered the value of $f$.
Assuming a constant $\Gamma$ in the region of highest wind acceleration, \citet{Grafener2017} derive an analytical expression connecting $\eta$ and $\tau_{F,\mathrm{s}}$ in terms of  $\varv_\infty/\varv_\mathrm{esc}$:
\begin{equation}
\begin{array}{c@{\qquad}c}
\eta \approx \dfrac{\tau_{F,\mathrm{s}}}{1+\dfrac{\varv_\mathrm{esc}^2}{\varv_\infty^2}}   \;\;\;\; \mathrm{or} \;\;\;\; \dfrac{\eta}{\tau_{F,\mathrm{s}}}\Big(\dfrac{\varv_\infty}{\varv_\mathrm{esc}} \Big) \approx \Big(1+\dfrac{\varv_\mathrm{esc}^2}{\varv_\infty^2}\Big)^{-1}
\end{array}
\label{eq: f_relation_graf}
\end{equation}
The above equation parameterizes the relation between $\eta$ and $\tau_{F,\mathrm{s}}$ in terms of the terminal-to-escape velocity ratio. 

How correctly does the formula estimate values of $f$? \citet{Vink2012} provide values of $f = \eta/(\tau_{F,\mathrm{s}} \approx 1)$ from their model atmosphere calculations at Galactic $Z$ for varying $\varv_\infty/\varv_\mathrm{esc}$ and $\beta$ (see their Table 1). For $\varv_\infty/\varv_\mathrm{esc} = 2.5$ and $\beta = 0.5$, they find $f$ of $\approx 0.83$. For a lower $\varv_\infty/\varv_\mathrm{esc} = 1.5$ and $\beta = 0.5$, they find $f \approx 0.68$. The values predicted from the formula for $\varv_\infty/\varv_\mathrm{esc} = 2.5$ and $1.5$ are $0.86$ and $0.69$ respectively, agreeing quite well with values of $f$ reported for $\beta$ of 0.5. However, for $\varv_\infty/\varv_\mathrm{esc} = 2.5$ and a higher $\beta$ of $1.5$ (typical for VMS), the value of $f$ reported in \citet{Vink2012} is $\approx 0.61$, while the formula still predicts $0.86$.

%\textbf{The velocity stratification obtained from our hydro models in the inner and outer parts can be separately fit using $\beta$-law functions. The inner quasi-static part is well approximated by $\beta \approx 2.4$ while the outer wind is best matched with $\beta \approx 1.5$ across all models.} 

%This is the value adopted in \citet{Sabhahit2022} to calculate the transition mass-loss rate in the Arches and 30 Dor clusters. 

Using our detailed hydro-models, we intend to obtain a similar functional relation as Eq. \ref{eq: f_relation_graf}, but more suitable for VMS. Using a straight line fit of the form $y = m\cdot x + c$ where $x = (1+\varv_\mathrm{esc}^2/\varv_\infty^2)^{-1}$, we get $m = 0.79(\pm 0.05)$ and $c = - 0.1(\pm 0.03)$. We take an approximate relation as follows:
\begin{equation}
\begin{array}{c@{\qquad}c}
\dfrac{\eta}{\tau_{F,\mathrm{s}}}\Big(\dfrac{\varv_\infty}{\varv_\mathrm{esc}} \Big)  \approx 0.75\Big(1+\dfrac{\varv_\mathrm{esc}^2}{\varv_\infty^2}\Big)^{-1}
\end{array}
\label{eq: f_powr_approx}
\end{equation}
The formula can now predict values of $f$ for a given $\varv_\infty/\varv_\mathrm{esc}$, accounting for the specific velocity fields of VMS which are reflected by higher $\beta$ values in empirical analyses. Plugging in  $\varv_\infty/\varv_\mathrm{esc} = 2.5$ at solar metallicity now gives $0.64$, agreeing with the value of $f$ reported for  $\varv_\infty/\varv_\mathrm{esc} = 2.5$ and higher $\beta = 1.5$ in \citet{Vink2012}. We use the formula in Eq. (\ref{eq: f_powr_approx}) to estimate values of wind efficiency parameter at sonic point optical depth $\tau_{F,\mathrm{s}}$ of unity.

\begin{figure*}
    \includegraphics[width = \textwidth]{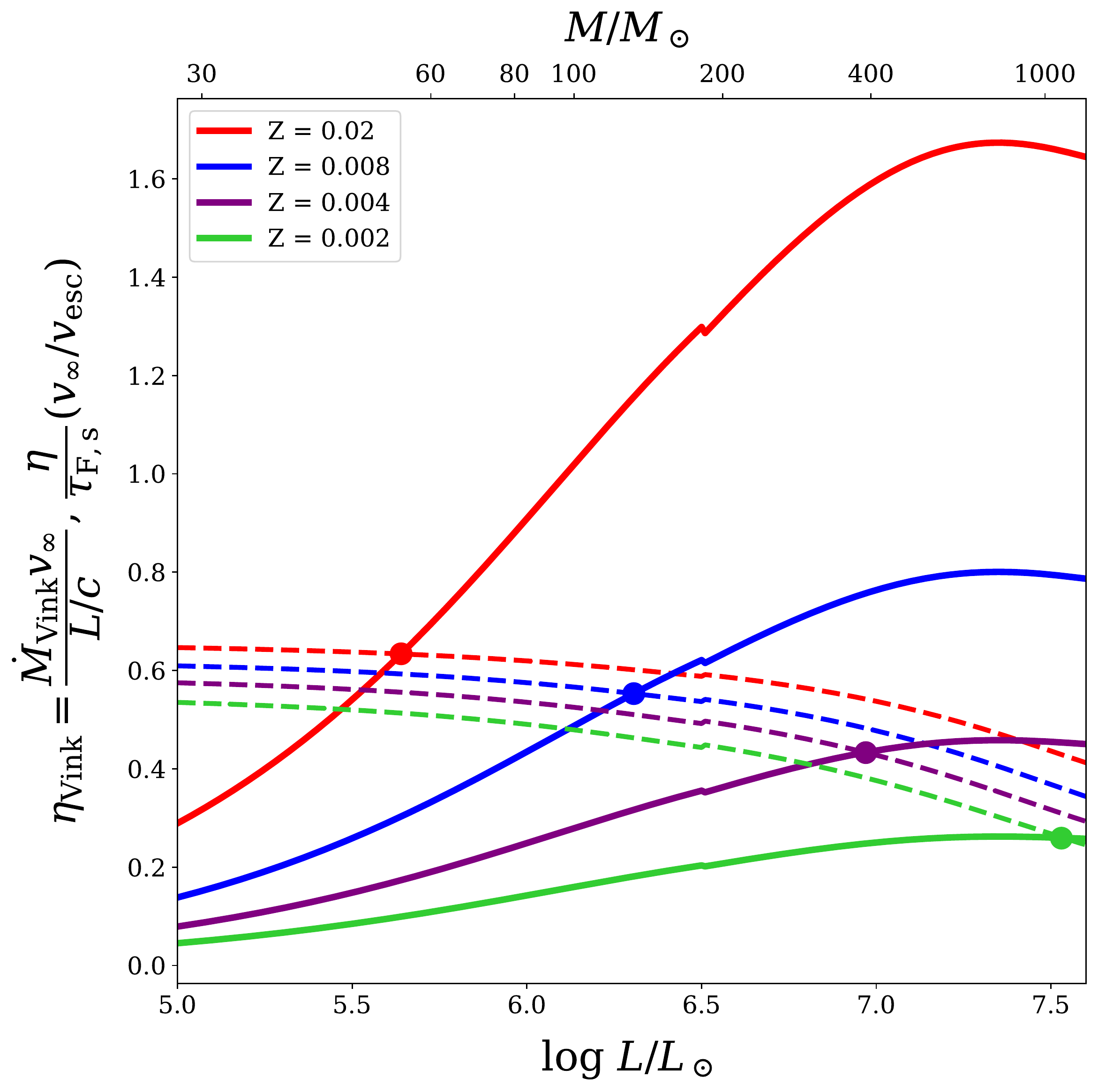}
    \caption{Variation of wind efficiency parameter as a function of luminosity (solid lines) using mass-loss rates from \citet{Vink2001} and terminal velocities from \citet{Lamers95}. The dashed lines show the variation of the parametric form of $f$. Different colors represent different metallicities ranging from $Z = 0.02$ to $0.002$. The location where a solid and dashed line cross denotes the transition on the zero-age main sequence at that corresponding metallicity. The small wiggle in the plots at log($L/L_\odot$) = 6.5 is due to switch in the mass-luminosity relation adopted from \citet{Graf2011}}.
    \label{fig:eta_L1}
\end{figure*}

\section{Transition properties with metallicity and temperature}
\label{sec: transition_prop_predict}

In this section, we perform a simple exercise to predict the variation of transition luminosity and mass loss with metallicity. These predictions can be compared with the `slash' stars in the Arches and 30 Dor.  We use the absolute rates from \citet{Vink2001} in this section, and the arguments will be based on the mass-loss scaling as predicted by these MC models.

Stars begin their MS with a uniform chemical composition, i.e., the surface and central abundances are the same. Given a sequence of luminosity $L$ (say $10^5-10^{8} \; L_\odot$ ) and a fixed surface hydrogen $X_\mathrm{s} = X_\mathrm{c} = 0.7$, the corresponding masses $M_\mathrm{hom}$ can be obtained using the mass-luminosity relations under homogeneity assumptions  \citep{Graf2011} (two relations are used here to cover the entire range of masses involved). We disregard the small changes in initial H content with metallicity which will be considered in detailed evolution models. With the mass and luminosity known, and fixing the metal content, one can use the mass-loss relations from \citet[their Eq.\,24]{Vink2001} to estimate an absolute mass-loss rate. 
MC models updated for dynamical consistency above 60 $M_\odot$ show almost no variation of mass-loss rate with effective temperature \citep[see][]{Vink2011, VS2021}, and so we do not include a temperature scaling. The quantity $\dot{M}_\mathrm{Vink}$ henceforth is the mass-loss rate predicted from \citet{Vink2001} ignoring the $T_{\mathrm{eff}}$ terms. 

Additionally, the terminal velocity is calculated from \citet{Lamers95} as $\varv_\infty = 2.6\varv_\mathrm{esc, eff}$, where $\varv_\mathrm{esc, eff}$ is the effective escape velocity  corrected for the classical Eddington parameter, i.e., $M_\mathrm{eff} = M_\mathrm{hom}(1-\Gamma_\mathrm{e})$. There is a radius parameter in the effective escape velocity and ideally, one uses the critical radius (cf. Eq. \ref{eq: v_esc}). But this quantity is not known a priori from hydro-static structure models. For the simple analysis here, we take the radius estimate from the Stefan-Boltzmann law connecting the luminosity and an assumed $T_\mathrm{eff}(R_{\tau_\mathrm{Ross} = 2/3})$. For stellar evolution models later, we use the temperature from the outer boundary radius which is also defined at $\tau_\mathrm{Ross} = 2/3$. 

As for the variation of $\varv_\mathrm{\infty}$ with $Z$, a weak scaling of $0.20$ is taken. The terminal velocities of OB stars in the ULYSSES sample hint at a weak $Z$ dependence of $0.22(\pm 0.03)$ \citep{Hawcroft2023}.  Such a shallow scaling with $Z$ is further corroborated by the wind models from \citet{Leitherer1992} as well as MC calculations with updated dynamical consistency \citep{VS2021}. 

\citet{Lamers95} measure terminal velocities of 117 stars with spectral type O through F and report an observed bi-stability in the terminal velocities at spectral type B1. Across this Bi-stability Jump (BSJ), the terminal velocities are lower by a factor of two and the terminal velocity is $1.3$ times the effective escape velocity. We also include a bi-stability in the mass loss \citep{Vink99} in our models. Note that this has been disputed by hydro-dynamically consistent radiation-driven wind modelling with the \textsc{fastwind} code \citep{Bjorklund2021} that does not predict a mass-loss bi-stability jump\footnote{Note that while \cite{Krticka21} also predict lower mass-loss rates than those of \cite{Vink2000}, they do recover the mass-loss increase at the bi-stability jump.} However, the predicted wind velocities from that modelling are an order of magnitude larger than observed. Given that dynamically consistent MC modelling \citep{Vink2018,VS2021} do predict terminal wind velocities in the correct ballpark, for now we take this as a good reason for including the mass-loss increase at the bi-stability jump.

With both mass loss and terminal velocity estimated, the wind efficiency parameter is calculated. We call this quantity $\eta_\mathrm{Vink}$ as it uses $\dot{M}_\mathrm{Vink}$,
\begin{equation}
\begin{array}{c@{\qquad}c}
\eta_\mathrm{Vink} \coloneqq \eta(\dot{M}_\mathrm{Vink}) = \dfrac{\dot{M}_\mathrm{Vink}\varv_\infty}{L/c}
\end{array}
\label{eq: eta_vink}
\end{equation}
Each individual value of luminosity $L$ gives a value of mass $M_\mathrm{hom}$, an estimate of mass-loss rate $\dot{M}_\mathrm{Vink}$, an escape velocity $\varv_\mathrm{esc}$, an effective escape velocity $\varv_\mathrm{esc, eff}$, a terminal speed $\varv_{\infty}$ and a wind efficiency parameter $\eta_\mathrm{Vink}$.

In Fig.\,\ref{fig:eta_L1} we show the variation of $\eta_\mathrm{Vink}$ (solid lines) as a function of luminosity. The corresponding homogeneous masses are provided on the upper x-axis. The same exercise is performed at four different metallicities: $Z = 0.02$ (GAL, red), 0.008 (LMC, blue), 0.004 (SMC, purple) and $Z = 0.002$ (tenth, green). For a fixed $L$, there is a steep decrease in absolute values of $\eta_\mathrm{Vink}$ with metallicity, as both mass loss and terminal velocities decrease with $Z$. 

For a fixed $Z$, the quantity $\eta_\mathrm{Vink}$ increases with luminosity (and mass), and only slightly dips at extremely high luminosities above log($L/L_\odot$) $\approx 7.2$. The dip is due to the terminal velocity term which depends on the product $M\cdot(1-\Gamma_\mathrm{e})$. This product is non-monotonous with $L$, and effectively lowers the terminal velocities towards higher luminosities. 

To determine where the switch to an optically thick wind occurs, we make use of the parametric form of $\eta/\tau_{F,\mathrm{s}}$ from Eq. (\ref{eq: f_powr_approx}). The dashed lines denote the wind efficiency numbers required to drive an enhanced wind at each luminosity. In the luminosity range log($L/L_\odot) \approx 5-6.5$, the dashed lines are almost horizontal with very little variation with $L$. This is because $\varv_\infty/\varv_\mathrm{esc, eff}$ is constant with luminosity and the values of $\Gamma_\mathrm{e}(L)$ are not high enough to significantly affect $\varv_\infty/\varv_\mathrm{esc}$ and consequently $\eta/\tau_{F,\mathrm{s}}$. As the luminosity increases, there is a noticeable drop in the dashed lines, due to the ratio  $\varv_\infty/\varv_\mathrm{esc}$ now weakly scaling inversely with $L$. The weak inverse scaling with $L$ is also in qualitative agreement with our PoWR models (cf.\ the last column of the first three models in Table \ref{table:powr_tau1}).

The location where the \textit{cross-over} between the solid and dashed lines occurs is where we define our `switch' to an enhanced wind in MESA. The relevant equation to obtain the location of the colored dots is:
\begin{equation}
\begin{array}{c@{\qquad}c}
\eta_\mathrm{Vink}(L) = \dfrac{\eta}{\tau_{F,\mathrm{s}}}\Big(\dfrac{\varv_\infty}{\varv_\mathrm{esc}} \Big) 
\end{array}
\label{eq: eq1}
\end{equation}
\begin{equation}
\begin{array}{c@{\qquad}c}
\varv_\infty = 2.6 \sqrt{\dfrac{2GM_\mathrm{hom}(1-\Gamma_\mathrm{e}(L))}{R}}\Bigg(\dfrac{Z}{Z_\odot}\Bigg)^{0.20}
\end{array}
\label{eq: eq2}
\end{equation}
\begin{equation}
\begin{array}{c@{\qquad}c}
\dfrac{\eta}{\tau_{F,\mathrm{s}}}\Big(\dfrac{\varv_\infty}{\varv_\mathrm{esc}} \Big) = 0.75\Big(1+\dfrac{\varv_\mathrm{esc}^2}{\varv_\infty^2}(L)\Big)^{-1} \\
\end{array}
\label{eq: eq3}
\end{equation}

The first equation ensures that the wind efficiency parameter $\eta_\mathrm{Vink}$ equals $\eta/\tau_{F,\mathrm{s}}(\varv_\infty/\varv_\mathrm{esc})$ at the location of the switch. We need to solve for the luminosity at the cross-over point, which we call $L_\mathrm{switch}$, and the corresponding mass $M_\mathrm{switch,hom}$ that satisfies the first equation. The second equation provides the terminal velocity to evaluate $\eta_\mathrm{Vink}$. Metallicity enters the equation here. The third equation is the parametric form of $\eta/\tau_{F,\mathrm{s}}$ where the terminal-to-escape velocity ratio is estimated for each $L$ using Eq. (\ref{eq: eq2}).

\begin{table*}
\begin{tabular}{c c c c c c c c} 
\hline
$Z$ & $T_\mathrm{eff}(R_{\tau_\mathrm{Ross} = 2/3})$ kK & $X_\mathrm{s}$  & log($L_\mathrm{switch}/L_\odot$) & $M_\mathrm{switch, hom}/M_\odot$ & $\varv_\mathrm{\infty, switch}$(km/s) & log($\dot{M}_{\mathrm{switch}}$) & $\eta_\mathrm{switch}$  \\
\hline\hline
0.02 & 45 & 0.7  & 5.64 & 54.42 & 3220.92 & $-5.75$ & 0.634 \\
0.008 & 45 & 0.7  & 6.306 & 128.27 & 2421.7 & $-5.02$ & 0.553 \\
0.004 & 45 & 0.7  & 6.97 & 391.91 & 2014.92 & $-4.389$ & 0.433 \\
0.002 & 45 & 0.7 & 7.53 & 1109.96 & 1523.05 & $-3.92$  & 0.259 \\
\hline
0.02 & 30 & 0.7  & 6.023 & 86.45 & 2017.23 & $-5.18$ & 0.618 \\
0.02 & 20 & 0.7  & 5.718 & 59.44 & 706.69 & $-5.19$ & 0.427 \\
0.02 & 15 & 0.7 & 5.965 & 80.23 & 508.98 & $-4.82$ & 0.41 \\
\hline
\end{tabular}
\caption{The absolute values of $L_\mathrm{switch}$, the corresponding homogeneous masses $M_\mathrm{switch, hom}$ and the mass-loss rate at switch $\dot{M}_{\mathrm{switch}}$ for a given metallicity, effective temperature and $X_\mathrm{s} = 0.7$. The top row changes $Z$ for a constant $T_\mathrm{eff}$, while the bottom row changes $T_\mathrm{eff}$ at constant $Z$. } 
\label{tab: theoretical_f_values}
\end{table*}

\begin{table*}
\begin{tabular}{c c c c c c c c c} 
        \hline

        & $L_\mathrm{trans}/L_\odot$ & $T_\mathrm{eff}(R_{\tau_\mathrm{Ross} = 2/3})$ (kK) &  $X_\mathrm{s}$ & $\varv_\mathrm{\infty, trans} (\mathrm{km/s})$ & $M/M_\odot$ &  log($\dot{M}_\mathrm{D = 10}$) & log($\dot{M}_\mathrm{trans, f = 0.6}$)    \\
        \hline\hline
        Arches & $10^{6.06}$ & $33.9$ & $0.7$ & 2000 &  $90.76$  & $-5.19$ & $-5.16$\\
        30 Dor &  $10^{6.31}$ & $44.4$  &  $0.62$ & 2550 & $129.04$  & $-4.98$ & $-5.01$\\

        \hline
\end{tabular}
\caption{Average transition parameters of the `slash' stars from the Arches cluster in the Galaxy and the 30 Dor cluster in the LMC. The empirical luminosities, effective temperatures, terminal velocities and $X_\mathrm{s}$ are provided, and the corresponding homogeneous masses are obtained using mass-luminosity relations from \citet{Graf2011}. The ratio of $\varv_\infty/\varv_\mathrm{esc}$ can then be obtained by using the stellar mass and radius. Two different estimates for the mass-loss rates of these objects are also provided (see text for details).} 
\label{tab: empirical_transition}
\end{table*}

Fixing $T_\mathrm{eff} = 45 $ kK, $X_\mathrm{s} = 0.7$, the three equations can be satisfied to obtain $L_\mathrm{switch}$ and  $M_\mathrm{switch, hom}$ as a function of metallicity. The mass-loss rate at the location of the switch is then given by $\dot{M}_\mathrm{switch} = \dot{M}_\mathrm{Vink}(L_\mathrm{switch}, M_\mathrm{switch, hom})$. The wind efficiency parameter at the switch can be defined as
  \begin{equation}
   \begin{array}{c@{\qquad}c}
    \eta_\mathrm{switch} \coloneqq \eta_\mathrm{Vink}(L_\mathrm{switch}) = \eta/\tau_{F,\mathrm{s}}(\varv_\infty/\varv_\mathrm{esc}) 
    \end{array}
   \label{eq: eta_equality}
   \end{equation}
In Table \ref{tab: theoretical_f_values}, we list these quantities for different $Z$. The first three columns provide the initial assumptions for $Z$, $T_{\mathrm{eff}}$ and $X_\mathrm{s}$.  We would like to distinguish between the `transition' properties (mainly $\dot{M}_\mathrm{trans}$ and $L_\mathrm{trans}$) discussed in Sec. \ref{sec: transition mass loss} and the luminosity and mass loss obtained at the switch location discussed here. The transition mass loss in Sec. \ref{sec: transition mass loss} is derived from purely analytical arguments and uses empirical luminosities $L_\mathrm{trans}$ and terminal velocities of `slash' stars in young clusters to obtain a mass-loss estimate, the transition mass loss $\dot{M}_\mathrm{trans}$. Here we use mass-loss predictions from MC calculations to predict the value of luminosity $L_\mathrm{switch}$ and mass $M_\mathrm{switch, hom}$ at the location where the wind efficiency parameter crosses a certain value. Indeed, theoretical input from atmosphere models and empirically determined ratio of terminal velocities to escape velocities have gone into determining these parameters. 

In Fig. \ref{fig:eta_L1}, we immediately notice that the colored dots shift to the right as metallicity decreases, signifying higher $L_\mathrm{switch}$ at lower $Z$. The mass-loss rate at the cross-over point also gets higher which seems counter-intuitive as mass-loss rates are predicted to reduce towards lower $Z$. However, $L_\mathrm{switch}$ is higher and dominates the behavior of mass loss.  The variation of $\eta_\mathrm{switch}$ with metallicity depends on both the adopted scaling of the terminal velocity with $Z$ and the drop in $\eta/\tau_{F,\mathrm{s}}$ towards higher $L$. 
%We find a shallower dependence of $\eta_\mathrm{switch}$ with metallicity when compared to a direct extrapolation of the VMS recipe from S22 (cf. $\eta$ values in Fig. \ref{fig:S22_tot}).  

\begin{figure}
    \includegraphics[width = \columnwidth]{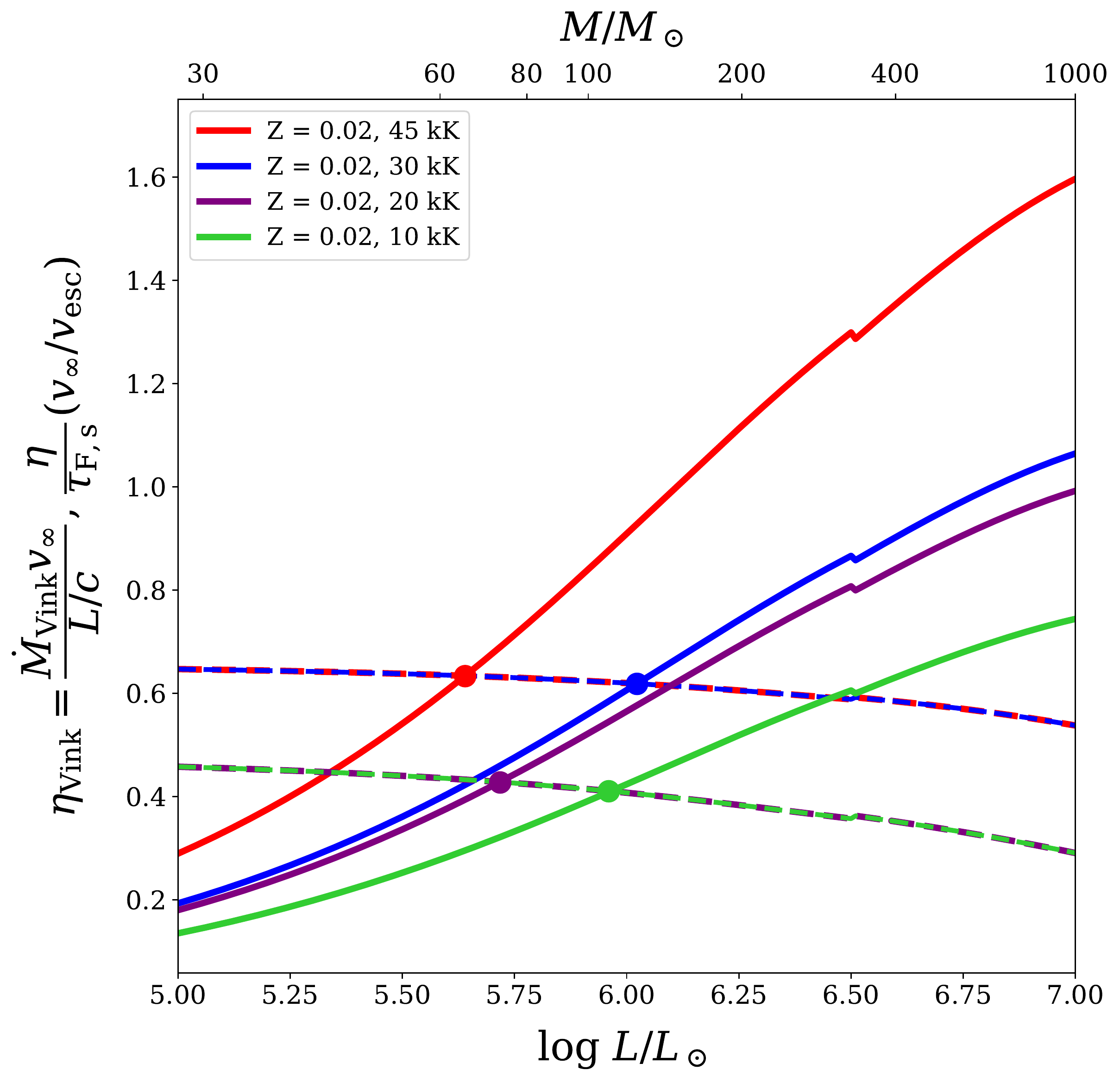}
    \caption{Same as Fig. \ref{fig:eta_L1}, but now testing the effect of temperature at constant $Z$. }
    \label{fig:eta_L2}
\end{figure}

Additionally, we test how temperature affects the cross-over point. We fix the metallicity and change $T_\mathrm{eff}$ to $30, \;20$ kK and $15$ kK.  Fig. \ref{fig:eta_L2} shows the variation of $\eta_\mathrm{Vink}$ and $\eta/\tau_{F,\mathrm{s}}$($\varv_\infty/\varv_\mathrm{esc}$) for four different effective temperatures. We pick two temperatures on the hot side of the  Bi-stability jump: 45 (red) and 30 kK (blue) and two on the cool side: 20 (purple) and 15 kK (green). 

On the hot side of the BSJ and for a constant $L$, $\eta_\mathrm{Vink}$ decreases towards lower temperatures. As the radius increases, the terminal velocity drops. When crossing the first BSJ at about $25$ kK, there is a jump in wind efficiency by approximately a factor of two. But more importantly, the quantity $\eta/\tau_{F,\mathrm{s}}$ is shifted downwards due to a lower terminal-to-effective escape velocity ratio on the cool side of the BSJ. This means it is easier to switch to an optically-thick wind on the cooler side of the BSJ.

How do the quantities in Table \ref{tab: theoretical_f_values} compare to the observed properties of `slash' stars in the Arches and 30 Dor? There are two such stars in the Arches and six in 30 Dor. These VMSs in different metallicity environments with different temperatures can help test our predictions. In Table \ref{tab: empirical_transition}, we provide `average' luminosities, terminal velocities, effective temperatures and corresponding homogeneous masses of the `slash' stars. We provide empirically determined mass-loss rate of the transition stars with micro-clumping factor $D_\mathrm{cl} = 10$ from \citet{Martins2008} and \citep{Best2014}. The transition mass loss estimate (Eq. \ref{eq: transition_mdot_third_form}) with $f = 0.6$ is also provided. 

The luminosity of the `slash' stars in the 30 Dor is higher compared to the Arches, and we correctly predict the trend of higher transition luminosity at lower $Z$. We also correctly predict the absolute values of the transition luminosities in both clusters when adopting the right temperatures (see the second and fifth row in Table \ref{tab: theoretical_f_values}). The exercise performed here with $X_\mathrm{s} = 0.7$ roughly corresponds to the zero-age main sequence (ZAMS), and successfully explains the observed transition properties in the Arches and 30 Dor. The mass-loss rate predicted from \citet{Vink2001} at the cross-over point agrees well with two independent calculations of the mass-loss rate of these objects.

Moreover, \citetalias{Sabhahit2022} assumes $f = 0.6$ irrespective of the host $Z$ when calculating the transition mass-loss rate. At Galactic (Z = 0.02) and LMC (Z = 0.008) metallicity, the values of $\eta_\mathrm{switch}$ obtained are 0.62 and 0.55 respectively, a shallow scaling with $Z$. The assumption of $f = 0.6$ to estimate the transition mass loss in the Arches and 30 Dor is reasonable. 

In summary, there is a remarkable agreement between the observed transition properties in the Arches and 30 Dor clusters, and the properties we predict at the switch. This gives confidence in our new approach which can be extended to study the evolution of VMS at low $Z$.

\section{Towards a VMS mass-loss framework at lower Z}
\label{sec: vms recipe Z}

\begin{figure}
    \includegraphics[width = \columnwidth]{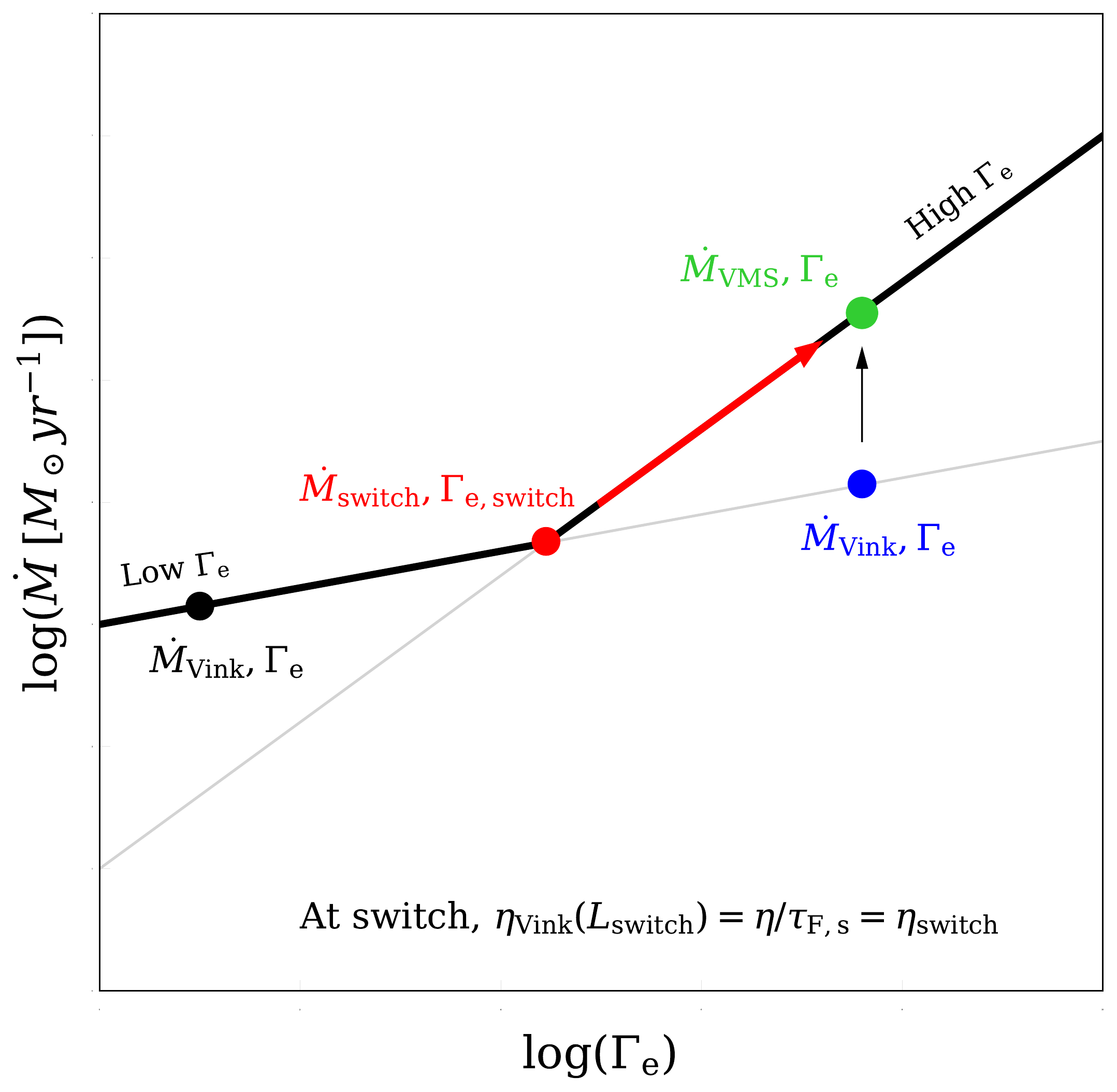}
    \caption{Schematic representation of the new VMS mass-loss framework where the condition to switch to a high-$\Gamma_\mathrm{e}$, optically-thick wind is based on wind efficiency parameters. A step-by-step method is described in the text. }
    \label{fig:cartoon}
\end{figure}

In this section, we would like to briefly describe the VMS mass-loss implementation in \citetalias{Sabhahit2022} and the general framework to extend the rates to lower metallicities (Fig. \ref{fig:cartoon}).

The VMS mass-loss implementation in \citetalias{Sabhahit2022} has two parts. It consists of an \textit{optically-thin} wind recipe suitable for canonical O stars with \textit{low values} of $\Gamma_\mathrm{e}(<0.4)$. This part has a shallow mass-loss scaling with stellar luminosity ($+2.194$) and mass ($-1.313$) based on MC models from \citet{Vink2001}. The second part consists of an optically-thick wind with a much steeper mass-loss scaling with luminosity ($+4.77$) and mass ($-3.99$) taken from \textit{high}-$\Gamma_\mathrm{e}$ models of \citet{Vink2011}. The switch condition to an enhanced wind in \citetalias{Sabhahit2022} is based on the maximum of the two absolute mass-loss rates. The logic behind such an implementation is that below the transition point, the low-$\Gamma_\mathrm{e}$ part is chosen as it predicts higher absolute rates. The implementation is intended to smoothly switch when the luminosity becomes high enough above which the high-$\Gamma_\mathrm{e}$ part starts predicting a higher mass-loss rate.

\citetalias{Sabhahit2022} also assessed the relative importance of electron number density compared to the ratio $L/M$ on the mass loss and argued that VMS models using a mass-loss scaling where $L/M$ dominates over $X_\mathrm{s}$ evolve at almost constant temperatures by dropping their luminosity. The drop in luminosity struck a balance between the enhancement in the mass loss and envelope inflation, resulting in models evolving vertically downwards in the HRD. Such a vertical evolution provides a simpler and more natural explanation for the narrow range of observed VMS temperatures in the two clusters. 

As for the absolute mass-loss rate, \citetalias{Sabhahit2022} uses the aforementioned concept of transition mass loss in the Arches and 30 Dor clusters to anchor the \textit{base rates} for the optically-thick part of the recipe.  This requires the availability of young clusters massive enough to host multiple O and WNh stars along with transition objects to obtain a transition luminosity and terminal velocity. The lack of such young massive clusters below LMC metallicity hinders this approach of determining the transition mass-loss rate at lower $Z$.

The new approach is based on wind efficiency numbers and seeks to address both the aforementioned limitations. Here we provide a step-by-step description of our new mass-loss framework and also highlight certain improvements.

Let $L$, $M$ and $X_\mathrm{s}$ be the model-predicted stellar luminosity, mass and surface H mass fraction at time $t$, which gives the electron scattering Eddington parameter $\Gamma_\mathrm{e}$ of the model. The effective temperature, surface H and metallicity required to determine the switch properties are taken from detailed stellar evolution models. The simple exercise in Sec. \ref{sec: transition_prop_predict} is now performed iteratively as the model is evolving.  The switch luminosity is obtained using the condition $\eta_\mathrm{Vink}(L_\mathrm{switch}, M_\mathrm{switch,hom}) = \eta/\tau_{F,\mathrm{s}}(\varv_\infty/\varv_\mathrm{esc})$. This gives the quantity $\Gamma_\mathrm{e,switch}$. As the model is evolving, if $\Gamma_\mathrm{e}$ crosses $\Gamma_\mathrm{e,switch}$, we have reached the transition and we stop evaluating the switch luminosity. Above this point, we switch to an optically-thick wind. Similar to \citetalias{Sabhahit2022}, the new method also has two components as seen in Fig. \ref{fig:cartoon}. 

\setenumerate[1]{wide = 0pt,labelwidth = 0.5cm, leftmargin =!}
\begin{enumerate}
  \item For the low-$\Gamma_\mathrm{e}$ regime where $\Gamma_\mathrm{e} < \Gamma_\mathrm{e,switch}$, the absolute mass-loss rate is
  \begin{equation}
\begin{array}{c@{\qquad}c}
\dot{M}_\mathrm{low\; \Gamma_\mathrm{e}} = \dot{M}_\mathrm{Vink}
\end{array}
\label{eq: V01_mass_loss}
\end{equation}
  In Fig. \ref{fig:cartoon}, we represent this by a black dot at $\Gamma_\mathrm{e}$ and $\dot{M}_\mathrm{Vink}$. This lies to the left of the `switch', represented by a red dot. This condition ensures that below a certain $\eta$, the mass-loss rates always correspond to an optically-thin wind. This includes the bi-stable nature predicted in their wind near $25$ kK, where the mass-loss rates are increased by a factor of approximately 5 due to changes in the ionization balance of Fe. The terminal velocities drop with an overall increase in $\eta$. The bi-stability temperature is calculated as the model is evolving and depends on the metallicity: $T_\mathrm{Bistability\; jump}(\mathrm{in \;kK}) = 61.2 + 2.59\cdot(-13.64 + 0.89\; \mathrm{log}(Z/Z_\odot))$ \citep{Vink99}.  The absolute rates predicted by \citet{Vink2001} have been challenged in the classical O star regime, where the uncertainties in the mass loss can be up to a factor of 2-3. \citep{Bouret2003, Ft2006}. These uncertainties are mostly due to the clumping in the wind and the treatment of radiative transfer in the sub-sonic region. There is also the so-called `weak-wind problem' where empirical estimates of mass-loss rates for late-type O-dwarfs are significantly lower, up to a magnitude, than expected from theoretical predictions. However, \citet{Vink2012} have shown that \citet{Vink2001} correctly predicts the mass-loss rate close to the observed transition. 
  \item If $\Gamma_\mathrm{e} > \Gamma_\mathrm{e,switch}$ at any point during the evolution, we have entered the high-$\Gamma_\mathrm{e}$ regime. In this case, the mass loss as predicted by \citet{Vink2001} is represented by a blue dot, and still lies on the shallow-slope line but to the right of the red dot. A steeper $\Gamma_\mathrm{e}$ scaling from \citet{Vink2011} is now applied starting from the red dot until the green dot is reached  (which has the same $L$ and $M$ as the blue dot). Mathematically, this is given by 
  
\begin{equation}
\begin{array}{c@{\qquad}c}
\dot{M}_\mathrm{high\; \Gamma_\mathrm{e}} = \dot{M}_\mathrm{switch}\times\Bigg(\dfrac{L}{L_\mathrm{switch}}\Bigg)^{4.77} \Bigg(\dfrac{M}{M_\mathrm{switch}}\Bigg)^{-3.99}
\end{array}
\label{eq: VMS_mass_loss}
\end{equation}
Since both $\Gamma_\mathrm{e}$ of the model and $\Gamma_\mathrm{e}$ at the switch are calculated iteratively, the term $1+X_\mathrm{s}$ can be cancelled. The high-$\Gamma_\mathrm{e}$ part forms the second component of our implementation and is essentially a `boost' applied to the mass-loss rates from \citet{Vink2001}, with a steeper scaling of mass loss with mass and luminosity.  For now, we retain both the $Z$-scaling from \citet{Vink2001} when evaluating $\dot{M}_\mathrm{Vink}(L_\mathrm{switch}, M_\mathrm{switch})$, but investigate the effects of having a weaker $Z$-scaling on VMS in Sec. \ref{sec: diff_t_z_scaling}.

\end{enumerate}

To summarize, the location of the switch is based on wind efficiency parameters while the amount of mass loss `boost' is set by the difference between the model $\Gamma_\mathrm{e}$ and the $\Gamma_\mathrm{e}$ at the switch.
Further checks are performed when switching from one wind recipe to another. For example, the switch to the high-$\Gamma_\mathrm{e}$ wind is performed only if $\Gamma_\mathrm{e}$ is increasing with time. Similarly, a switch back to the low-$\Gamma_\mathrm{e}$ part is carried out only if $\Gamma_\mathrm{e}$ is decreasing with time.

The new mass-loss framework is more robust and solves most of the problems raised earlier.  The issue of vastly different $\eta$ at the location of the switch for different initial masses is automatically resolved as the quantity $\eta_\mathrm{switch}$ remains relatively unchanged during the evolution. But the most important advantage is that this framework can now be extended to lower $Z$ even if VMS are not available for empirical determination of the transition mass loss.  A step-by-step procedure to implement the above VMS mass loss framework is provided in Appendix \ref{Appendix: framework_logic}.

The new framework, however, has two main caveats. First, the method determines $M_\mathrm{switch,hom}$ for a given $L_\mathrm{switch}$ under the assumption of chemical homogeneity at the location of the switch. The homogeneous mass $M_\mathrm{switch,hom}$ is the \textit{maximum} possible mass for that given $L_\mathrm{switch}$. This can lead to an over-prediction of  the VMS mass loss as evaluated by Eq. (\ref{eq: VMS_mass_loss}). Note that the quantity $M_\mathrm{switch}$ appears in the numerator of Eq. (\ref{eq: VMS_mass_loss}), and any over-prediction in the mass directly affects our VMS mass loss. 

VMS with initial mass above 160 - 200 $M_\odot$, or log $(L/L_\odot)$ above $\approx 6.5$, remain almost fully mixed throughout the MS \citep{Yusof2013, Kohler2015, Sabhahit2022}. The $L_\mathrm{switch}$ values listed in table \ref{tab: theoretical_f_values} (first four rows) suggest that the chemical homogeneity assumption used in the new framework breaks down at Galactic metallicity.  At Galactic $Z$, the implementation in \citetalias{Sabhahit2022} might be more appropriate as the absolute rates are anchored based on the empirical results from the Arches cluster. For the case of $Z = 0.008$ (LMC-like), $L_\mathrm{switch}$ is slightly below the threshold luminosity for the chemical homogeneity assumption. Consequently, our new work over-predicts the mass loss, but the difference between using the new framework and the implementation in S22 should be small (about 0.1 dex difference in absolute mass loss). The caveat discussed here should be resolved for values of $Z$ below 0.008 as $L_\mathrm{switch}$ increases into the regime where the homogeneity assumptions hold well.

Second, the framework is only as good as the mass loss and terminal velocity estimate used to build it. The crux of this method relies on the condition $\eta_\mathrm{Vink} = \eta/\tau_{F,\mathrm{s}}(\varv_\infty/\varv_\mathrm{esc})$ at the switch. The quantity $\eta_\mathrm{Vink}$ uses a mass loss estimate from \citet{Vink2001} and terminal velocities from \citet{Lamers95}. Any over-prediction or under-prediction by the Vink rates or the terminal velocity will affect our framework.

%This prevents us from having a single expression to calculate mass-loss rates from an optically-thick wind at different masses and metallicities. We provide an 'average over time' absolute mass-loss rate for each metallicity in Appendix \ref{}, but would caution its usage to systematically study VMS evolution.

%\citet{Vink2012} obtained a wider range of values for $f \approx 0.4-0.8$ at Galactic metallicity, with $f$ increasing with increasing $\varv_\infty/\varv_\mathrm{esc}$ and decreasing $\beta$. Using  $\varv_\infty/\varv_\mathrm{esc} = 2.5$ and most appropriate value for $\beta = 1.5$ \citep{Vink2011}, they arrived at $f\approx 0.6$. 

%Our Galactic models predict slightly lower values of $f\approx 0.48$. Combining the weak metallicity scaling of $\varv_\infty/\varv_\mathrm{esc}$ with Eq. \ref{eq: f_relation_graf}, we find the factor $f$ to also weakly scale with $Z$. So we fix the value of $f$ to $\approx 0.6$ for all metallicities. 

% effective beta value is about 1.8

%Despite these uncertainties, the absolute rates predicted by \citet{Vink2001} matches the transition mass-loss rate derived in the Arches Cluster. In Table \ref{Tab: mdot_ccomp} we list the mass-loss rate predicted for the transition objects using different approaches:

\begin{figure*}
    \includegraphics[width = \textwidth]{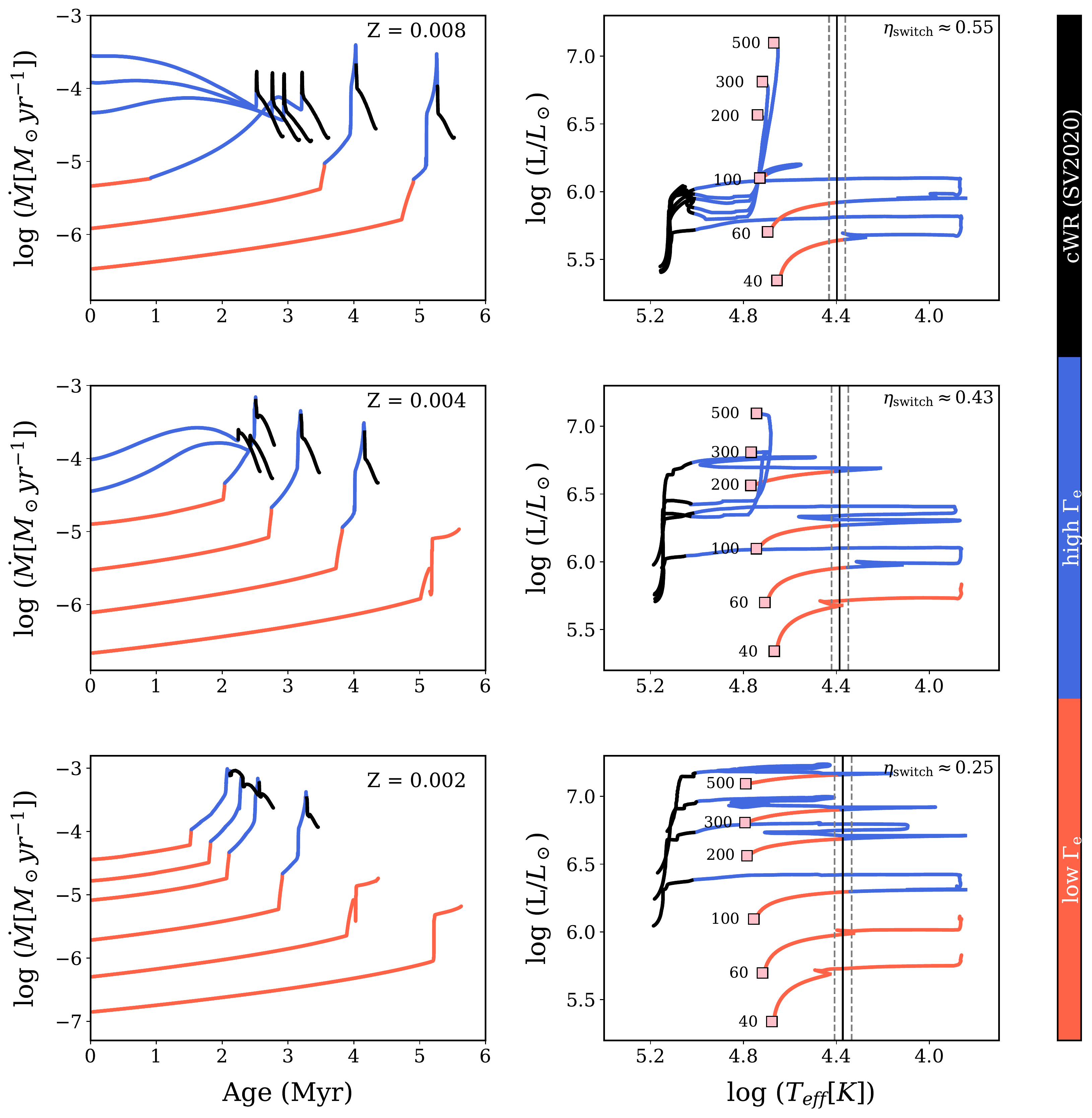}
    \caption{Absolute mass-loss rates predicted by the new mass-loss framework during the Main Sequence (left), and the corresponding HRD tracks (right). Only selected models from the grid are shown here.  The initial masses shown are 40, 60, 100, 200, 300 and 500 $M_\odot$. Three different initial metallicities are considered: $Z = 0.008, 0.004$ and $0.002$ roughly corresponding to LMC, SMC and tenth solar metallicity. The plot is color-coded based on the mass loss used. The vertical black line marks the bistability jump temperature for each $Z$. A temperature range of $dT = 1000 \;K$ (grey dashed lines) is used to smoothly transition from the hot side of the jump to the cool side.}
    \label{fig: HRD_tracks_new_models}
\end{figure*}

\section{MESA modelling}
\label{sec: mesa}

In this section, we give a general overview of the different parameter inputs to model the evolution of VMS. We detail the evolution code used and the treatment of different physical processes such as overshooting and rotational mixing.

The VMS models presented in this paper are calculated using the 1D stellar evolution code MESA (version r12115) \citep{MESA11, MESA13, MESA15, MESA17, MESA19}. The relevant mixing and rotation parameters are as follows: mixing length parameter $\alpha_\mathrm{MLT} = 1.5$, an exponential overshooting prescription from \citet{Herwig2000} above the H core with efficiency $f_\mathrm{ov} = 0.03$ and a semiconvective diffusive efficiency of $\alpha_\mathrm{sc} = 1$. The models in the full grid are non-rotating, but we investigate the role played by rotation at high $\Gamma$ using an initial $\Omega_\mathrm{  ZAMS}/\Omega_\mathrm{crit}$ of $0.6$.  The critical $\Omega$ is defined as $\Omega_\mathrm{crit}^2 = (1-L/L_\mathrm{Edd})GM/R^3$, where $L_\mathrm{Edd} = 4\pi Gc M/\varkappa_\mathrm{Ross}$, is the Eddington luminosity \citep{MESA13}. In the rotating model, the diffusion coefficients for different rotation-induced instabilities follow \citet{Heger2000}. 

The initial masses considered in this study are $40,$ $60,$ $80,$ $100,$ $120,$ $150,$ $180,$ $200,$ $250,$ $300,$ $400$ and $500\; M_\odot$, The initial metal mass fraction values are $Z_\mathrm{tot} = 0.008, 0.004, 0.002, 0.001$ and $0.0002$. The metallicity values considered here should approximately span from LMC to a hundredth solar metallicity. The new implementation is not used at Galactic metallicity as it over-predicts the mass-loss rates, and we refer to \citetalias{Sabhahit2022} for Galactic models. The grid consists of $12\times5$ models (12 different initial masses, 5 different metallicities).

 All models begin with the following chemical composition distributed uniformly throughout the star. The initial He mass fraction $Y$ in our models is calculated as $Y = Y_{\text{prim}} + (\Delta Y/\Delta Z) Z$ where the primordial He abundance $Y_{\text{prim}} = 0.24$ and $(\Delta Y/\Delta Z) = 2$. The hydrogen mass fraction $X$ is then given by $X = 1-Y-Z$. Both  $Y_{\text{prim}}$ and $(\Delta Y/\Delta Z)$ follow the default values in MESA \citep{Pols1998}. During the pre-MS, a constant accretion rate of dlogM/step = $5\times10^{-3}$ is used until the initial mass is reached. The model is then allowed to relax until it arrives on the zero-age main sequence. The evolution is followed till the end of core He burning where the central He mass fraction falls below $Y_\mathrm{c} < 0.01$.

During the MS, we use the mass-loss framework detailed in the previous section. Detailed wind models accounting for opacities from  different elements show that Fe is the dominant wind driver in the radiation-driven winds of massive stars. As surface iron does not change during the MS evolution of our models, we scale the mass loss with the initial $Z$ instead of the total surface $Z$.

The models are further run till the end of core He burning.  The mass loss used in the core He-burning phase is:
\begin{enumerate}
 \item For temperatures between $4,000$ K and $100,000$ K, we utilize the same mass-loss implementation used for the MS. There is no inherent reason for the mass-loss mechanism near the surface to depend explicitly on the type of fuel being burnt in the core. Moreover, cool LBVs observed at temperatures below $\sim 10,000$ K show particularly strong winds believed to be due to their proximity to the Eddington limit. So we extend our high-$\Gamma_\mathrm{e}$ mass loss beyond the MS. 
 \item Below $T_\mathrm{eff} \approx 4,000 $K we switch to the cool red supergiant recipe by \citet{deJager1988}.  Following the arguments from \citet{Vink2021}, the blue supergiants in the temperature range $8\,000-12\,000\,$K are too hot to form dust and their winds are rather expected to be driven by iron-dominated opacities.
\item if $T_\mathrm{eff} > 100,000$ K and $X_\mathrm{c}$ falls below 0.01, we use the classical WR mass-loss recipe from \citet{SV2020} which is employed in the He burning phase of the evolution.
 
\end{enumerate}

Certain models encounter convergence issues during the core He-burning phase of the evolution. To retain consistency and ensure models complete the evolution, we use the MLT++ option throughout the core He burning phase of the evolution.

\subsection{HRD evolution}
\label{sec: MS_evo}

In this section, we present non-rotating models, their HRD tracks and the evolution of mass loss as a function of time. In Fig. \ref{fig: HRD_tracks_new_models} we show the absolute mass-loss rate (left) and the corresponding evolutionary tracks (right) for three different metallicities. Two different colors are used to show the two components of our mass loss, red implementing the low-$\Gamma_\mathrm{e}$ mass loss from \citet{Vink2001} suitable near the transition and blue implementing the high-$\Gamma_\mathrm{e}$ mass loss as outlined before.  The classical WR mass-loss recipe from \citet{SV2020} is indicated in black. In the top right corner of the HRDs, we mention the approximate value of the wind efficiency parameter where the switch occurs to the high-$\Gamma_\mathrm{e}$ wind. This value of $\eta$ corresponds to the flux-weighted mean optical depth $\tau_{F,\mathrm{s}}$ crossing unity and approximately coincides with the spectral morphological transition from O to WNh stars. 
%We begin by directly comparing the model results obtained with our new approach with the VMS models published in S22 (Fig. \ref{fig:S22_tot}).

The two fundamental parameters affecting the mass-loss rate in our models are $L/M$ ratio and initial $Z$. During the MS evolution of the star, luminosity increases due to the increase in the mean molecular weight $\mu$ \citep{Kipp1990, Farell2021}. Simultaneously the total mass of the star decreases owing to mass loss. Both contribute to increasing $L/M$. A significant decrease in the total mass can potentially reduce the total luminosity output of the star. Which effect dominates -- the $\mu$-effect or the mass effect -- then depends on how strong the mass loss is.

Models using the low-$\Gamma_\mathrm{e}$ wind evolve across the HRD to cooler temperatures.  Mass loss is not strong enough to cause a drop in the total luminosity. However, as the initial mass increases, and consequently the $L/M$ ratio, our models gradually shift towards using the high-$\Gamma_\mathrm{e}$ part of the implementation. These models can evaporate a considerable fraction of their initial mass leading to an overall drop in luminosity. Whether stars evolve horizontally or vertically downwards depends on the mass-loss rate used during the majority of the MS.

Certain models switch to the high-$\Gamma_\mathrm{e}$ wind just after evolving off the ZAMS.  These models mark the `boundary' between the two distinct evolutionary behaviors. For example, consider the 100 $M_\odot$ initial mass model at $Z = 0.008$ on the ZAMS. The $L/M$ is not high enough to warrant the usage of the high-$\Gamma_\mathrm{e}$ wind. But as the star evolves off the MS there is an increase in $L$ and a simultaneous decrease in $M$, and above a certain $L/M$ the model enters the high-$\Gamma_\mathrm{e}$ regime. Any further increase in $L/M$ results in a stronger mass loss due to the adopted scaling, causing the star to strip its envelope and expose its deeper hotter layers.

As $Z$ decreases, this `boundary' between the two distinct evolutionary behaviors shifts towards higher initial masses. This is in line with the results found earlier, that both $L_\mathrm{switch}$ and $M_\mathrm{switch}$ shift upwards at lower $Z$. 
For $Z = 0.004$ (middle row), only the highest mass models switch to the VMS mass loss on the hot side of the BSJ. 
However, the exact model that first starts using this enhanced wind depends on the  $Z$-scaling adopted in our VMS implementation. For example, if the VMS $\dot{M}$-$Z$ scaling is shallower than the predictions for OB stars (of $\approx 0.85$), then it would be easier to switch to an enhanced wind already at a lower initial mass. We explore this further in Sec. \ref{sec: diff_t_z_scaling}.

 %They form a deep convective zone capable of dredging up burning products to the surface. 

Models below the transition typically begin their core-helium burning phase as YHGs with temperatures below $10,000$ K.  Multiple LBVs are observed at such cool temperatures which come in two different flavors - "quiescent" LBVs that lie along the S-Dor instability strip with moderate winds which have been explained using radiation pressure on spectral lines and are likely $Z$-dependent \citep[][]{Vink2002, Smith2004, Grassitelli2021} and the constant temperature outburst phase near $10,000$ K, such as displayed by Eta Car in 1840 which might be continuum-driven and relevant even at low $Z$ \citep{Smith2006, Van2008}.  We collectively call the objects in the temperature range between $4,000$ and $10,000$ the YHG/LBV regime \citep[see also][]{Smith2004, Koumpia2020}. The physics of LBV-type mass loss is still very uncertain, but both observations and wind models point towards strong mass loss due to the proximity to the Eddington limit. We automatically switch to the high-$\Gamma_\mathrm{e}$ wind at these temperatures as $\eta_\mathrm{switch}$ is lowered across the bi-stability jump.
 
\begin{figure}
    \includegraphics[width = \columnwidth]{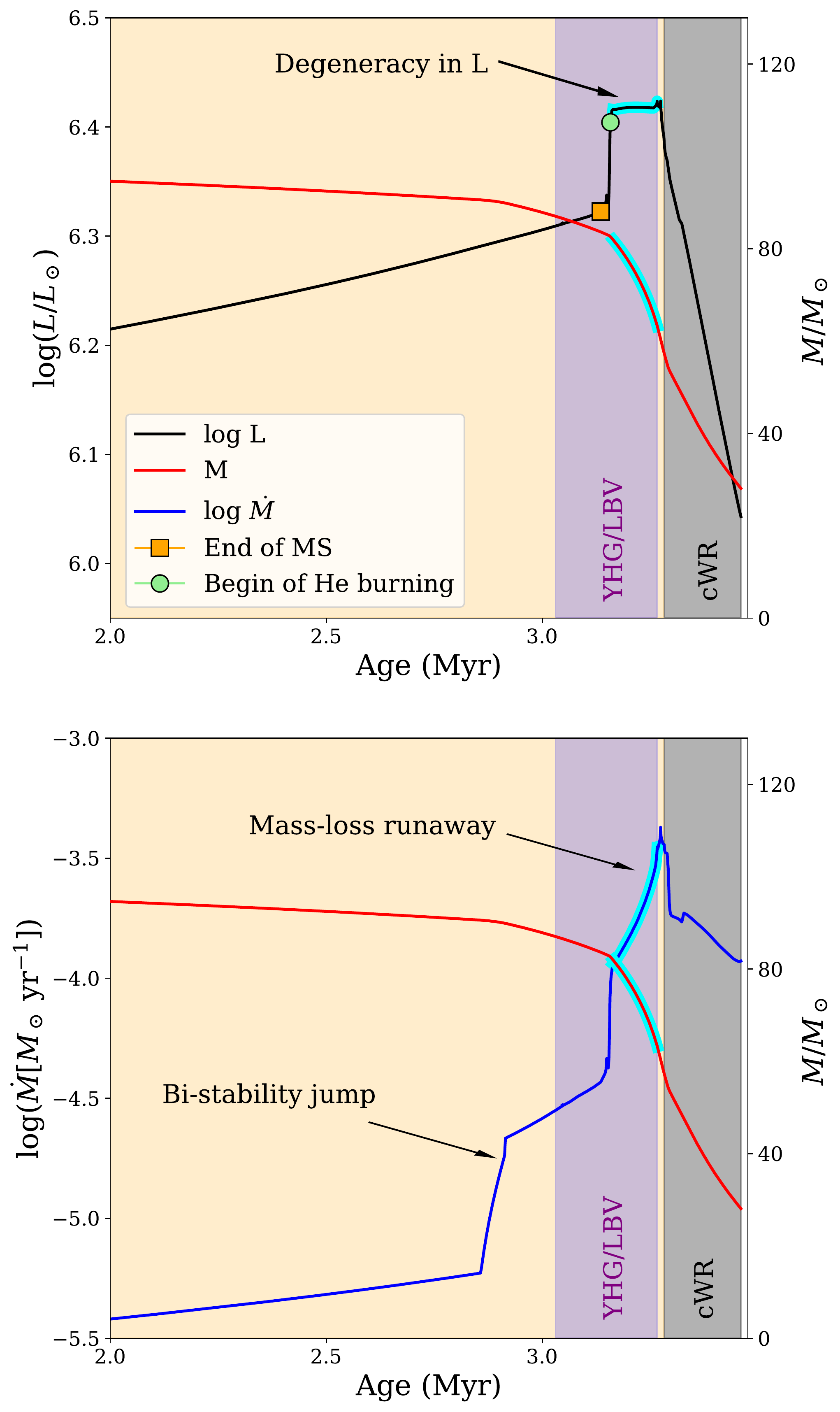}
   \caption{(Top) Evolution of luminosity and total mass of a 100 $M_\odot$ model at $Z = 0.002$. The purple-shaded region indicates the YHG phase with surface temperatures below 10,000 K. The grey-shaded region marks the classical WR phase with temperatures above 100,000 K. The orange-shaded region marks the rest of the temperatures.}
    \label{fig:mass_loss_runaway}
\end{figure}

Although we do not explicitly include an LBV mass-loss prescription in our models, we still mimic high mass loss. In Fig. \ref{fig:mass_loss_runaway} (top), we plot the evolution of luminosity (black line) and total mass (red line) of a 100 $M_\odot$ VMS model at $Z = 0.002$. The corresponding mass-loss evolution (blue line) is shown in the bottom plot. Once the model settles on the core He burning phase, we notice a degeneracy in the total luminosity marked by the cyan region. Despite the ever-decreasing total mass, the luminosity during this part of the evolution remains almost a constant \citep[see also snapshot models of ][]{Farrell2020}.

Since mass loss scales inversely with total mass, a decrease in total mass increases the mass loss which further decreases the total mass, ultimately resulting in a mass-loss runaway which ends once the entire envelope is stripped. The accelerated increase of the mass loss in the bottom plot highlights the mass-loss runaway effect.
While this runaway effect in our models is significantly more pronounced due to the jump in mass loss across the BSJ, it can still occur without the BSJ albeit very weakly.

High values of surface $\Gamma_\mathrm{e}$, in excess of 0.8, are recorded during the runaway phase. Appropriately, the mass loss is completely dictated by the $\Gamma_\mathrm{e}$ parameter, which dominates over $Z$. The mass-loss runaway can never be reproduced using the red-supergiant mass-loss recipe by \citet{deJager1988} as there is no explicit dependence of mass loss on the total mass. This behavior is unique to models that employ a $L/M$-scaling in the YHG/LBV regime.

\begin{figure*}
    \includegraphics[width = \textwidth]{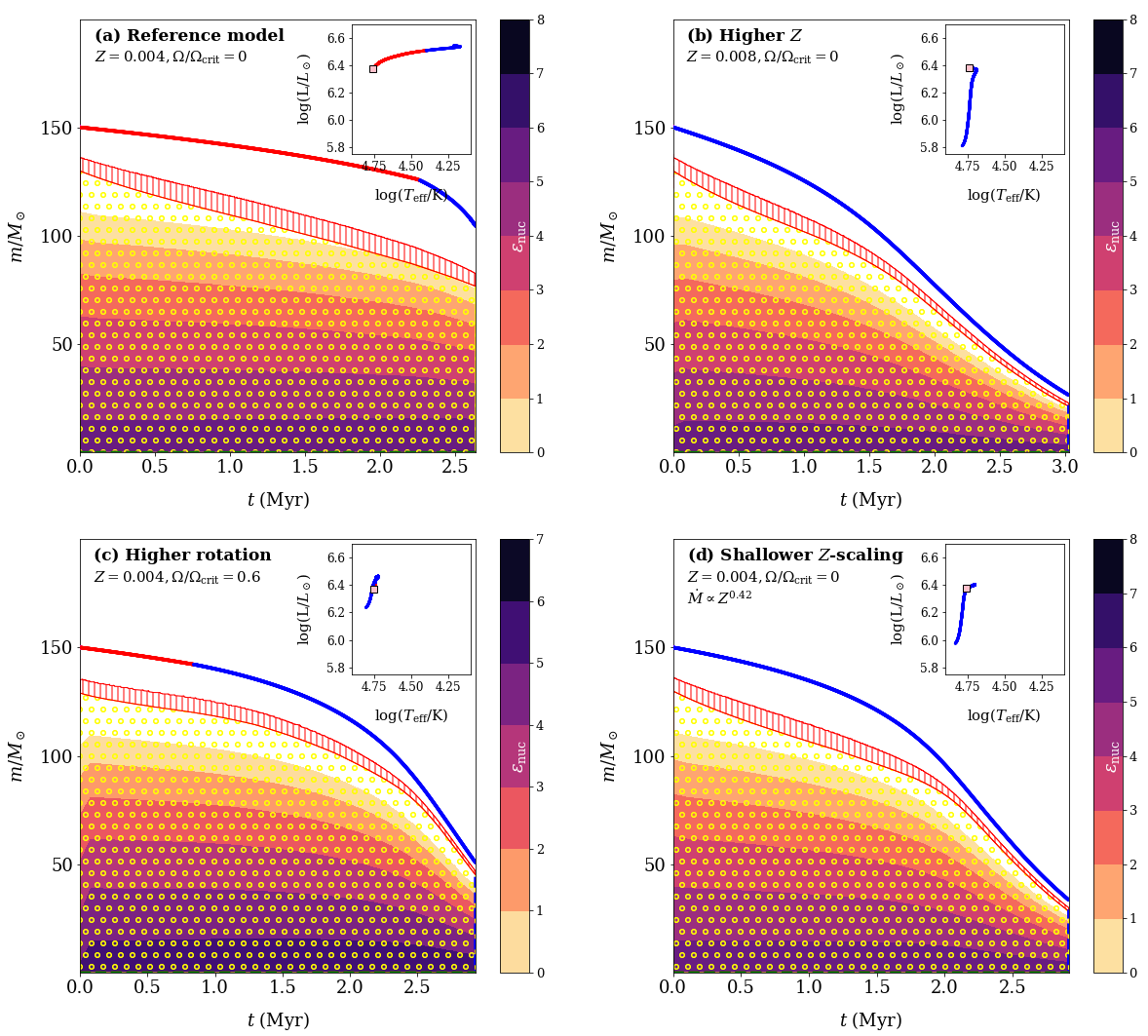}
    \caption{Kippenhahn plots showing the internal profile evolution of 150 $M_\odot$ VMS models. Model (a) is the reference model with metallicity $Z = 0.004$ and non-rotating. Model (b) increases the metallicity to $Z = 0.008$. Model (c) is a rotating model with $Z = 0.004$ and initial rotation of $\Omega_\mathrm{ZAMS}/\Omega_\mathrm{crit} = 0.6$. Model (d) has the same input parameters as the reference model but with a shallower $Z$-scaling for the VMS mass loss.  In the inset, we show the HRD track of the corresponding model. The yellow-dotted region shows convection while red shaded region denotes the overshooting region above the convective core. For the total mass evolution, we use the same color code as in Fig. \ref{fig: HRD_tracks_new_models} depending on the mass loss used.   }  
    \label{fig: kipp_compare}
\end{figure*}

From the beginning of core He burning until the model reaches $\sim 100,000$ K, the luminosity remains almost constant and the mass loss continuously increases. Eventually, the convective envelope is stripped and the star begins evolving bluewards. Once the surface effective temperature crosses $100,000$ K, we switch to the WR recipe from \citet{SV2020}. The model has lost all of its H envelope and is a pure He star. The luminosity once again becomes sensitive to the total mass and we see an overall drop in the total luminosity due to the high WR mass loss. In contrast, models that evolved vertically downwards during the MS employing the enhanced wind are almost fully mixed by the end of the MS. These models quickly lose their thin envelope and become pure-He stars without becoming a YHG. The final He core masses and the consequences of the new VMS mass loss on the PISNe boundary are discussed in Sec. \ref{sec: PISN}.

\subsection{Effect of rotation}
\label{sec: rot_effects}

To showcase the effect of metallicity on the evolution of interiors of VMS, we compare Kippenhahn diagrams of non-rotating models at different $Z$.  In Fig. \ref{fig: kipp_compare}(a), we track the internal profile evolution of the 150 $M_\odot$ model at $Z = 0.004$.  This model acts like the reference model for the discussions below. In Fig. \ref{fig: kipp_compare}(b), we show the evolution of a 150 $M_\odot$ model with initial $Z = 0.008$. At higher $Z$, the star is above the transition and uses the high-$\Gamma_\mathrm{e}$ wind during its entire evolution. It loses $\approx 75\%$ of its initial mass already during the MS. The enhanced mass loss results in chemically homogeneous evolution throughout the MS. On the other hand, the $Z = 0.004$ model only switches to the high-$\Gamma_\mathrm{e}$ wind on the cool side of the BSJ. This difference in mass loss directly affects the total mass and the size of the He core at the TAMS. This difference in total mass already during the MS can have ramifications for the final mass of the remnant formed.

Previous studies have examined the effects of rotation on mass loss \citep[for e.g.][]{Langer1998, MMevo2000}. Radiation-driven wind models from \citet{FA1986} accounting for rotation found an increase in the mass-loss rate. Rotating stellar evolution models often boost the mass-loss rate using a fit to the results in \citet{FA1986}: $\dot{M} = \dot{M}_\mathrm{vrot = 0}(1-\Omega/\Omega_\mathrm{crit}$)$^{-0.43}
$ \citep[see for example][]{Heger2000}. However, there is still debate about whether rotation increases the mass-loss rate for O stars. For example, \citet{MV2014} use MC models with dynamical consistency extended to the non-spherical 2D case and find the (surface-averaged) total mass-loss rates to be lower than the spherical 1D case. 

However, when close to the radiative Eddington limit, \citet{MM2000} demonstrated the existence of the so-called $\Omega\Gamma$-limit while accounting for the von Zeipel theorem and the existence of two critical velocities. Above a certain $\Gamma (> 0.639)$, the $\Omega\Gamma$-limit can be reached even for non-extreme rotations and the break-up near the equator is reached sooner for rotating models. 

Here, we explore how rotation affects stars near the transition. To this end, we use the following approximate formula from \citet{MM2000} to evaluate the mass-loss rate for the rotating model:
\begin{equation}
\begin{array}{c@{\qquad}c}
\dfrac{\dot{M}(\Omega)}{\dot{M}(\Omega = 0)} \approx \dfrac{(1-\Gamma)^{\frac{1}{\alpha}-1}}{\Big(1-\frac{4}{9} (\frac{\varv}{\varv_\mathrm{crit, 1}})^2 -\Gamma\Big)^{\frac{1}{\alpha}-1}}
\end{array}
\label{eq: rotation_mass_loss}
\end{equation}
The critical velocity is given by $\varv_\mathrm{crit,1} = \Big(\frac{2}{3}\frac{GM}{R_\mathrm{pb}}\Big)^{0.5}$ where $R_\mathrm{pb}$ is the polar radius at break-up. We further assume that the polar radius does not change with rotation, i.e., $R_\mathrm{pb}/R_\mathrm{p}(\Omega) = 1$.

Rotation affects our mass-loss formalism in two different ways. First, the exercise performed in Sec. \ref{sec: transition_prop_predict} uses a mass-loss estimate from \citet{Vink2001} to obtain the switch luminosity. An increase in $\dot{M}_\mathrm{Vink}$ results in a lower $L_\mathrm{switch}$. Or in other words, rotation can enable the transition to already occur for lower initial masses. Second, the non-zero rotational velocity in the presence of high values of $\Gamma$ means the mass-loss rates are higher due to the proximity to the $\Omega\Gamma$-limit according to Eq. (\ref{eq: rotation_mass_loss}).

We run a 150 $M_\odot$ initial mass model at $Z = 0.004$ that begins on the ZAMS with $\Omega/\Omega_\mathrm{crit} = 0.6$. We show the Kippenhahn evolution and the corresponding HRD track in Fig. \ref{fig: kipp_compare}(c).  At the TAMS the difference in total mass is $\approx$ 50 $M_\odot$. Similar to the non-rotating reference model, the evolution begins with the model being below the transition. However, due to the two effects of rotation mentioned above, a switch occurs to the high-$\Gamma_\mathrm{e}$ wind already on the hot side of the BSJ. The rotating model evolves  almost fully mixed on the MS owing to the higher mass loss in the proximity of the $\Omega\Gamma$-limit. 

 Due to the mass loss scaling with $Z$, one might expect the rotation to significantly affect evolution at high $Z$. This would be true if the star could retain its rotational velocity for most of the MS. However at higher $Z$, mass loss removes all the angular momentum and brakes down the star very quickly. The largest effect of rotation is at low $Z$. This is because below a certain $Z$, the mass loss is low enough for the star to retain most of its angular momentum and rotation effects will be relevant for the entire MS.  To summarize, rotation can shift the transition luminosity downwards, and can significantly affect the final mass of the remnant formed especially at low $Z$.

 \begin{figure*}
    \includegraphics[width = \textwidth]{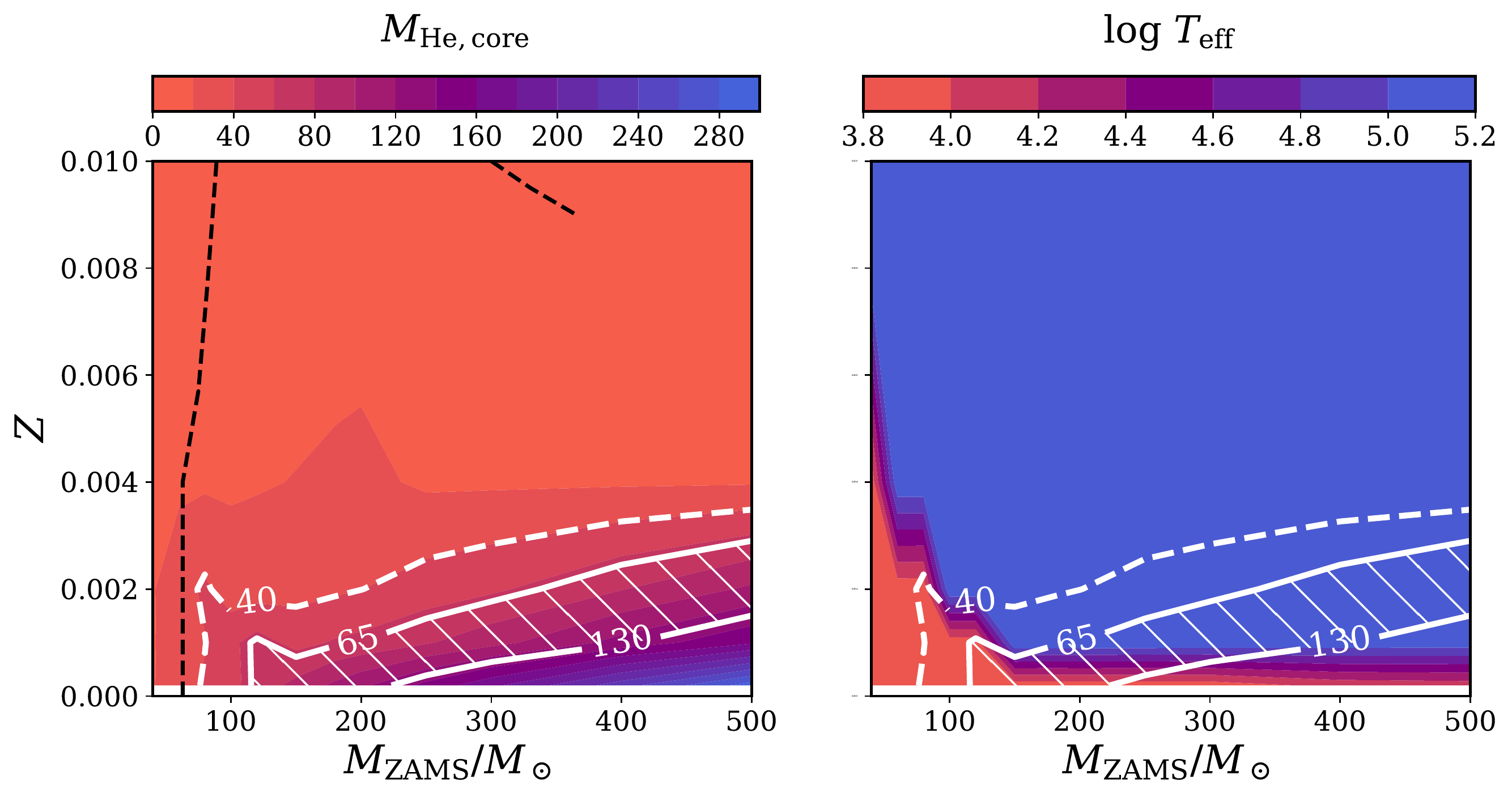}
   \caption{Contour plots showing He core mass (left) and effective temperature (right) at He core exhaustion as a function of initial mass and metallicity. The two solid white contour lines are for constant He core masses of 65 and 130 $M_\odot$, which mark the complete PISN boundary. The dashed white line marks the pulsational-PISN boundary at He core mass of 40 $M_\odot$ from our models.  For comparison, the pulsational-PISN boundary adapted from \citet{Spera2017} is shown in black dashed.}
    \label{fig: he_core_contour}
\end{figure*}

\subsection{Effect of $Z$-scaling of VMS mass loss}
\label{sec: diff_t_z_scaling}

The quantitative dependence of $\dot{M}$ on $Z$ has been theoretically investigated for decades \citep{CAK1975, Kudritzki1987}. On the observational side, mass-loss rates empirically derived from H$\alpha$ and ultraviolet (UV) wind lines for O and early B-type stars in the Galaxy, LMC and SMC show a decrease towards lower $Z$ \citet{Mokiem2007}.  

So far in this paper, we have used the $Z$-scaling derived from MC models in \citet{Vink2001}: $\dot{M} \propto \varv_\infty^{p} Z^{0.85}$ with $p = -1.226$.   Adopting a weak power-law scaling for the terminal velocities with metallicity, $\varv_\infty \propto Z^q$ where $q = 0.20$, the overall mass-loss dependence on metallicity works out to $\dot{M} \propto Z^{0.85 + pq} = Z^{0.6}$. These MC models use a $\beta$-law velocity stratification in the wind to obtain the mass-loss rate. 

Recently \citet{VS2021} used MC models updated for dynamical consistency and arrived at a shallower $Z$-scaling for mass loss of $\dot{M} \propto Z^{0.42}$ on the hot side of the BSJ \citep[see also][]{Alex2022}. We implement this shallower $Z$ dependence on a non-rotating, 150 $M_\odot$ initial mass model at $Z = 0.004$. We  plot the Kippenhahn evolution and HRD track in Fig. \ref{fig: kipp_compare}(d).  Owing to the shallow $Z$-scaling, both absolute mass loss and $\eta$ are higher. The switch to the high-$\Gamma_\mathrm{e}$ wind occurs earlier already on the hot side of the BSJ (cf. with the reference model). The model (d) spends its MS above the transition and evolves chemically homogeneously towards the end of the MS.

\section{Implications for the $Z$ threshold for PISN}
\label{sec: PISN}

One of the fundamental problems in  massive star evolution theory is to piece together the end fates of massive stars given the initial conditions such as initial mass and metallicity. Various studies in the past have tackled this question in the context of the role played by $Z$-dependent wind mass loss \citep[e.g.][]{Heger2003, Langer2007, Spera2017}. 

Stellar evolution theory suggests that VMS may undergo electron-positron pair creation leading to instabilities during core oxygen burning. Whether these objects end up directly collapsing into a BH or undergo pair instabilities that result in pulsations or a complete disruption of the star, depends on the core mass.  If pair instabilities in the core only give rise to pulsations, multiple mass loss episodes can occur and a BH forms. As our models do not follow the pulsations or the collapse, we adopt He core mass values often found in the literature - 65 $M_\odot$ and 130 $M_\odot$ - between which PISN is predicted to occur leaving behind no remnant \citep{HW2002, Heger2003}. These are the predicted final He core masses that map the boundaries for total pair-instability supernova. While stellar models at Galactic $Z$ show strong mass loss that is capable of preventing a PISN, the $Z$-dependence of mass loss means there exists a $Z$ threshold for the occurrence of PISN. For pulsational-PISN, the threshold is more uncertain with helium core masses likely between 37-45 $M_\odot$ \citep[for e.g.][]{Wooseley2017, Farmer2019}.  

Based on their evolution models, \citet{Langer2007} discussed the evolutionary status of the progenitors of PISN. Slow rotators produce YHGs with a $Z$ threshold for PISN of about $Z_\odot/3$. Fast rotators become classical WR stars with an extremely low $Z$ threshold for PISN of $\approx Z_\odot/1000$ due to strong winds in the WR regime. 

\citet{Spera2017} used the SEVN population synthesis code to study the BH mass spectrum as a function of $Z$ with evolution tracks for VMS taken from \citet{Chen2015}. \citet{Spera2017} mapped the end fates of VMS - direct collapse BH, pulsational and complete PISN - to their initial mass and metallicity (their Fig. 3). There is a clear `peak' in the parameter region that is predicted to suffer pair instabilities. We plot their pulsational-PISN boundary in Fig. \ref{fig: he_core_contour} (black dashed line) for the purpose of comparison. As we show below, this `peak' is heavily suppressed in our models due to the mass-loss runaway.

%The models in \citet{Chen2015} use the absolute mass-loss rates from \citet{Vink2011} including a kink above which mass-loss rates are enhanced. However, \citet{Vink2011} have noted their absolute mass-loss rates might be under-estimated. Rather than adopting the absolute mass-loss rates predicted by the MC models in \citet{Vink2011}, we have anchored the location of their kink using the concept of transition mass loss, and appropriately enhanced the mass loss above it.  

We would like to revisit the metallicity threshold for PISN in the context of VMS winds based on the concept of transition mass loss. Our models are calculated until core He exhaustion with the mass-loss framework discussed in this work applied to both core H and He burning phases. We assume negligible changes to the total and core masses in the last few thousand years during core C and O burning. In Fig. \ref{fig: he_core_contour} (left), we show a contour plot of final He core masses as a function of initial mass and initial $Z$. We draw solid contour lines  corresponding to constant He core masses of $65$ and $130 \; M_\odot$ marking the boundary for total PISN. Models within the shaded region are predicted to leave behind no remnant. The PPISN boundary is marked at helium core mass of 40 $M_\odot$ by a dashed contour line.

In the right figure, we show a contour plot of the surface temperature at He exhaustion. We superimpose the previously obtained contour lines of 40, 65 and 130 $M_\odot$ core masses. These two plots together can provide the status of our models only a few thousand years before the supposed pair-instability. The candidate progenitors for PISN range from YHGs to classical WR stars. Below $M_\mathrm{init} \lesssim 150 M_\odot$, the PISN progenitors are all YHGs while above $M_\mathrm{init} \gtrsim 300 M_\odot$, the progenitors are all classical WR stars. In between $150$ and $300 M_\odot$, the progenitor status depends on the initial metallicity.

Our non-rotating VMS models suggest that the metallicity threshold for total PISN from YHGs is about $Z_\odot/20$, possibly explaining the lack of unambiguous detection of PISN in the local Universe. The lower $Z$ threshold of $Z_\odot/20$ from our models, when compared to \citet{Langer2007}, is because of the runaway mass loss during the YHG regime predicted by our mass loss framework. The `peak' seen in the black dashed line  (cf. white dashed line from this work) - the PPISN boundary from \citet{Spera2017} - is heavily suppressed in our models, due to the strong mass loss that remains relevant even at low $Z$. Although we know LBVs have particularly strong winds and the extensive nebula surrounding these objects may suggest eruptive mass loss, the LBV phenomenon is still uncertain and as such the threshold number reported here may not be the final answer.

\section{Summary and conclusions}
\label{sec: conclusions}

In this paper, we use the concept of transition mass loss to study the evolution of very massive stars as a function of host metallicity ranging from LMC-like to $Z_\odot / 100$. This is an extension of the work in \citet{Sabhahit2022} towards lower $Z$ where individually observed VMS are not available for constraining absolute mass-loss rates.

\citet{Vink2012} derived a model-independent way to characterize the spectral morphological transition from O stars to the WNh sequence. Given a young cluster hosting multiple O and WNh stars, one can employ a simple condition that relates the wind efficiency number to the sonic point optical depth $\eta = f \tau_{F,\mathrm{s}}$ to accurately obtain the mass-loss rate of the transition Of/WNh stars. Using the Potsdam Wolf-Rayet (PoWR) atmosphere code to parameterize the relation between $\eta$ and $\tau_{F,\mathrm{s}}$ in terms of the terminal and escape velocity, we devise a simple exercise to investigate the variation of transition luminosity and transition mass loss with $Z$. We obtain the so-called `switch' or `cross-over' point based on  wind efficiency numbers. The switch luminosity $L_\mathrm{switch}$ and the mass-loss rate are then compared to properties of transition objects empirically determined in the Arches (Galaxy) and the 30 Dor (LMC) clusters.

We describe our new VMS mass-loss framework that can be extended to low $Z$ despite the lack of individually observed VMS beyond our Local Universe.  We run a grid of VMS models till the end of core He burning with the new mass-loss framework implemented in the MESA stellar evolution code. We summarize the main results from this study below
\begin{enumerate}
\item The quantity $L_\mathrm{switch}$ increases towards lower $Z$, in agreement with the higher transition luminosity in the 30 Dor compared to the Arches. The values of switch luminosity $L_\mathrm{switch}$ obtained for $Z = 0.02$ and $0.008$ agree well with the luminosity of the transition objects in the respective clusters. There is also a remarkable agreement between the absolute mass-loss estimate from \citet{Vink2001} and the empirically determined mass loss with $D = 10$ in both clusters. 
\item During core H burning, the increase in the mean molecular weight $\mu$ results in an increase of the total luminosity output for the star. Combined with a decrease in the total mass, it is natural for the luminosity-to-mass ratio $L/M$ of a massive star to increase as it  evolves off the MS. Whether VMS models evolve horizontally in the HRD towards cooler temperatures or vertically downwards towards lower luminosities depends on the mass loss. VMS winds can quickly evaporate a significant fraction of the initial mass and can dominate over the $\mu$-effect to result in an overall reduction in luminosity.  
\item In the yellow hypergiant regime, there exists a degeneracy in the luminosity despite varying total mass. With luminosity losing its sensitivity to mass, a runaway occurs where a decrease in $M$ increases mass loss, which further reduces $M$. This process occurs till all of the H-rich envelope is removed and the star presents itself as a pure He star. 
\item High surface values of $\Gamma_\mathrm{e}$ are achieved in the YHG regime even at low $Z$. Luminous blue variables in this regime show strong winds and their winds are thought to be $Z$-independent and continuum-driven. Although we do not specifically treat LBV-type mass loss in our models, the high values of $\Gamma_\mathrm{e}$ completely dominate the mass loss resulting in strong winds regardless of $Z$. Once the entire envelope is lost, luminosity becomes sensitive to mass once more. Strong Wolf-Rayet winds further drop the luminosity in our models towards the end of core He burning.
\item  Rotation adds an additional layer of complexity to already uncertain physics of VMS. There still exist uncertainties regarding rotation and mass loss, both code-specific limitations and the uncertain physics close to break-up. We perform preliminary tests on how rotation affects our mass-loss framework. Rotation can shift the transition luminosity downwards enabling VMS winds already at a lower initial mass. Using the rotation-enhanced mass-loss framework from \citet{MM2000}, we predict significant differences in the final mass of our VMS models due to close proximity to the $\Omega\Gamma$ limit. This gives a scatter in the final masses obtained based on rotation. 
\item The question of the $Z$ threshold for PISN is investigated in the context of VMS winds. The candidate progenitors for PISN range from YHGs to classical Wolf-Rayet stars.  Our models suggest a low metallicity threshold for PISN from YHG progenitors of about $Z_\odot/20$.  The strong winds in the vicinity of the Eddington limit combined with the mass-loss runaway effect can quickly strip the envelopes in our models and lower the core masses. However, the $Z$ threshold reported in this paper might be further shifted due to the still uncertain physics of LBV-type mass loss.
\end{enumerate}

\section*{Acknowledgements}

We thank the anonymous referee for constructive comments that helped improve the paper. We warmly thank the MESA developers for making their stellar evolution code publicly available. JSV and ERH are supported by STFC funding under grant number ST/V000233/1.
AACS is funded by the Deutsche Forschungsgemeinschaft (DFG - German Research Foundation) in the form of an Emmy Noether Research Group -- Project-ID 445674056 (SA4064/1-1, PI Sander). AACS further acknowledges support by funding from the Federal Ministry of Education and Research (BMBF) and the Baden-Württemberg Ministry of Science as part of the Excellence Strategy of the German Federal and State Governments.

%%%%%%%%%%%%%%%%%%%%%%%%%%%%%%%%%%%%%%%%%%%%%%%%%%
\section*{Data Availability}

The necessary files to reproduce the models used in this article are available on a GitHub repository: \url{https://github.com/Apophis-1/VMS_Paper2}

%%%%%%%%%%%%%%%%%%%% REFERENCES %%%%%%%%%%%%%%%%%%

\bibliographystyle{mnras}
\bibliography{References} % if your bibtex file is called example.bib

%%%%%%%%%%%%%%%%%%%%%%%%%%%%%%%%%%%%%%%%%%%%%%%%%%

%%%%%%%%%%%%%%%%% APPENDICES %%%%%%%%%%%%%%%%%%%%%

\appendix

\section{Composition details of PoWR models}
\label{Appendix: powr}

Here we list the composition details used in the PoWR model calculations. In table \ref{table: elements}, we list the metal mass fractions for different elements.

\begin{table}
\centering
\begin{tabular}{c c} 

        \hline
        Element & Mass fraction \\
        \hline\hline
         $X$ & > 0 \\
         $Y$ & $1- X - Z$ \\
         $Z_\mathrm{C}$ & 1 $\times 10^{-4}$ \;$Z/Z_\odot$ \\
         $Z_\mathrm{N}$ & 0.015 $Z/Z_\odot$  \\
         $Z_\mathrm{O}$ & 5.5 $\times 10^{-5}$ \;$Z/Z_\odot$ \\
         $Z_\mathrm{Ne}$ & 1.257 $\times 10^{-3}$ \;$Z/Z_\odot$ \\
         $Z_\mathrm{Si}$ & 6.7  $\times 10^{-4}$ \;$Z/Z_\odot$ \\
         $Z_\mathrm{P}$ & 5.8  $\times 10^{-6}$ \;$Z/Z_\odot$ \\
         $Z_\mathrm{S}$ & 3.1  $\times 10^{-4}$ \;$Z/Z_\odot$ \\
         $Z_\mathrm{Cl}$ & 8.2  $\times 10^{-6}$ \;$Z/Z_\odot$ \\
         $Z_\mathrm{Ar}$ & 7.34 $\times 10^{-5}$ \;$Z/Z_\odot$ \\
         $Z_\mathrm{K}$ & 3.13 $\times 10^{-6}$ \;$Z/Z_\odot$ \\
         $Z_\mathrm{Fe}$ & 1.6 $\times 10^{-3}$ \;$Z/Z_\odot$ \\

        \hline
\end{tabular}
\caption{Table showing the mass fractions of different elements used in our PoWR atmosphere models.} 
\label{table: elements}
\end{table}

%\begin{figure}
 %   \includegraphics[width = \columnwidth]{switch_prop.pdf}
  %  \caption{Evolution of $L_\mathrm{switch}$, $M_\mathrm{switch}$ and $\eta_\mathrm{switch}$ of a $300 M_\odot$ initial mass star at four different metallicities. These quantities are only saved from our models if the VMS mass loss is used (i.e., if $\eta_\mathrm{Vink} > \eta/\tau_\mathrm{F,s}$). The plot is colour-coded according to the effective temperature to indicate the side of the bistability jump the model is in.}
   % \label{fig: switch_prop}
%\end{figure}

%In Fig. \ref{fig: switch_prop}, we show the evolution of $L_\mathrm{switch}$ and $\eta_\mathrm{switch}$ for a 300 $M_\odot$ initial mass star, for four different metallicities. The plot is color-coded according to the surface temperature. The Galactic and LMC models at 300 $M_\odot$ quite easily enter the regime of VMS mass loss and remain `hot' throughout the MS. It is clear that there is very little variation in $L_\mathrm{switch}$ and $\eta_\mathrm{switch}$ at each metallicity, despite significant changes in total luminosity and temperature as the model evolves. A simpler implementation could just adopt the absolute values from Table \ref{tab: theoretical_f_values} to switch to an enhanced wind.

\section{Step-by-step logic for implementing the framework}
\label{Appendix: framework_logic}

Here we provide a step-by-step procedure to implement the new framework in other codes
\setenumerate[1]{wide = 0pt,labelwidth = 0.5cm, leftmargin =!}
\setenumerate[2]{labelwidth =0.5cm, align = left, leftmargin =0.5cm}
\begin{enumerate}
\item Evaluate model $\Gamma_\mathrm{e}$ from Eq. (\ref{eq: gamma_e}) using model $L$, $M$ and $X_\mathrm{s}$.
\item An iterative search is performed in this step to obtain $L_\mathrm{switch}$. Take a value of luminosity $L_\mathrm{iter}$ between $10^5$ $L_\odot$ and $10^8$ $L_\odot$. For each $L_\mathrm{iter}$, the following quantities are evaluated:
\begin{enumerate}
\item The homogeneous mass $M_\mathrm{hom, iter}(L_\mathrm{iter}, X_\mathrm{s,model})$ from \citet{Graf2011}. This also gives a $\Gamma_\mathrm{e, iter}$ from Eq. (\ref{eq: gamma_e}).
\item The radius $R_\mathrm{iter}(L_\mathrm{iter}, T_\mathrm{eff, model})$  using the Stefan-Boltzmann law.
\item  Escape velocity $\varv_\mathrm{esc,iter}$ using the above mass and radius. Effective escape velocity using mass, radius and  $\Gamma_\mathrm{e, iter}$.
\item Terminal velocity estimate using Eq. (\ref{eq: eq2}) where $Z$ is the surface metal mass fraction of the model.
\item A mass loss estimate from \citet{Vink2001} without the temperature terms.
\item The wind efficiency number $\eta_\mathrm{Vink, iter}$ from  Eq. (\ref{eq: eta_vink})
\item The quantity $\eta/\tau_{F,s}$ from Eq. (\ref{eq: eq3}) using the terminal and escape velocities evaluated above.
\end{enumerate}
Repeat the above steps until one finds the value of $L_\mathrm{iter}$ for which $\eta_\mathrm{Vink, iter}$ equals $\eta/\tau_{F,s}$ (i.e., satisfies Eq. \ref{eq: eq1}). This is the switch luminosity $L_\mathrm{switch}$. The corresponding mass is $M_\mathrm{switch, hom}$ and mass loss is $\dot{M}_\mathrm{switch}$. 
\item Evaluate $\Gamma_\mathrm{e,switch}$ from $L_\mathrm{switch}$, $M_\mathrm{switch, hom}$ and $X_\mathrm{s}$.
\item Compare model $\Gamma_\mathrm{e}$ and $\Gamma_\mathrm{e,switch}$. If model $\Gamma_\mathrm{e}$ is below $\Gamma_\mathrm{e,switch}$, no enhancement in mass loss (Eq. \ref{eq: V01_mass_loss}).
\item Once model $\Gamma_\mathrm{e}$ crosses $\Gamma_\mathrm{e,switch}$, the transition is reached and the evaluation of $L_\mathrm{switch}$ is stopped. Mass loss is enhanced according to Eq. (\ref{eq: VMS_mass_loss}).   
\end{enumerate}

The necessary files (inlists and \text{run\_stars}) to reproduce the models in MESA are available here: \url{https://github.com/Apophis-1/VMS_Paper2}

%%%%%%%%%%%%%%%%%%%%%%%%%%%%%%%%%%%%%%%%%%%%%%%%%%

% Don't change these lines
\bsp	% typesetting comment
\label{lastpage}
\end{document}